\documentclass[12pt, letterpaper]{article}
\usepackage{natbib}
\usepackage{amsmath, amssymb, amsfonts, amsthm, bbm, bm, mathrsfs, stmaryrd}
\usepackage{graphicx, subfigure, float, booktabs, longtable, siunitx, color}
\usepackage{algorithm, algpseudocode}
\usepackage{caption}
\usepackage{multirow, threeparttable, url, enumitem}
\usepackage{setspace}
\usepackage{fancyhdr}
\usepackage{appendix}
\usepackage{hyperref}
\usepackage{makecell}

\algdef{SE}[DOWHILE]{Do}{doWhile}{\algorithmicdo}[1]{\algorithmicwhile\ #1}%

\newcommand{\mbf}[1]{\ensuremath{\mathbf{#1}}}

   \def\bmD{\mbf D}

   \def\bmI{\mbf I}

   \def\bmQ{\mbf Q}  
     
    \def\bmS{\mbf S}

\newcommand{\bfv}[1]{\ensuremath{\boldsymbol{#1}}}

   \def\bD{\bfv D}

   \def\bG{\bfv G}

   \def\bO{\bfv O}

   \def\bW{\bfv W}  
   \def\bX{\bfv X}  
   \def\bY{\bfv Y}

\newcommand{\bfsym}[1]{\ensuremath{\boldsymbol{#1}}}

\def\bbeta{\bfsym \beta}
             
           \def\bDelta {\bfsym {\Delta}}

\def\bpsi{\bfsym {\psi}}

\definecolor{darkred}{RGB}{150,50,50}
\definecolor{brown}{RGB}{250,100,100}
\definecolor{green}{RGB}{000,150,100}
\definecolor{navy}{RGB}{000,000,100}

\newtheorem{theorem}{Theorem}
\newtheorem{remark}{Remark}
\def\trans{^{\scriptscriptstyle \sf T}}
\begin{document}
\title{Optimizing precision in stepped-wedge designs via machine learning and quadratic inference functions}   
\author{Liangbo Lyu and
Bingkai Wang$^*$\\
Department of Biostatistics, School of Public Health, University of Michigan}
\date{}
\maketitle
\begin{abstract}
Stepped-wedge designs are increasingly used in randomized experiments to accommodate logistical and ethical constraints by staggering treatment roll-out over time. Despite their popularity, existing analytical methods largely rely on parametric models with linear covariate adjustment and prespecified correlation structures, which may limit achievable precision in practice. We propose a new class of estimators for the causal average treatment effect in stepped-wedge designs that optimizes precision through flexible, machine-learning-based covariate adjustment to capture complex outcome-covariate relationships, together with quadratic inference functions to adaptively learn the correlation structure. We establish consistency and asymptotic normality under mild conditions requiring only $L_2$ convergence of nuisance estimators, even under model misspecification, and characterize when the estimator attains the minimal asymptotic variance. Moreover, we prove that the proposed estimator never reduces efficiency relative to an independence working correlation. The proposed method further accommodates treatment-effect heterogeneity across both exposure duration and calendar time. Finally, we demonstrate our methods through simulation studies and reanalyses of two empirical studies that differ substantially in research area and key design parameters.
\end{abstract}
\noindent{\bf Keywords}: Average treatment effect, causal inference, robust inference, staggered roll-out, time-varying treatment effect.
\section{Introduction}

In randomized trials, the stepped-wedge design refers to a staggered roll-out of a treatment at randomized time points, such that all units begin the study under control and eventually receive the treatment by its conclusion. This design is particularly attractive in settings where it is logistically infeasible to deploy the treatment to all units simultaneously, as in a parallel-arm design, or when there is a strong prior belief that the treatment is beneficial and withholding it from some units indefinitely is undesirable \citep{li2022stepped}. Because of these practical advantages, stepped-wedge designs have gained substantial popularity across a wide range of applied fields, including health sciences, implementation sciences, education, and social sciences \citep{mdege2011systematic, beard2015stepped, varghese2025systematic}.
In Section~\ref{sec: motivating-examples}, we introduce two motivating studies that both adopt stepped-wedge designs but differ markedly in key design features, including the unit of randomization, overall sample size, and the number of roll-out periods.

Traditional analyses of stepped-wedge designs typically rely on parametric regression models for covariate adjustment, yet such models may be too restrictive to deliver meaningful precision gains in practice. In the context of cluster-randomized trials, \cite{hussey2007design} first introduced mixed-effect models that linearly adjust for covariates while accounting for within-cluster correlation; these approaches have since become standard and widely applied \citep{li2021mixed, nevins2023adherence}. 
There is a rich literature of subsequent methodological developments, such as random effects specifications \citep{kasza2019inference, kasza2019impact, voldal2022model}, treatment-effect heterogeneity \citep{kenny2022analysis, maleyeff2022assessing, lee2025analysis}, and robustness to model misspecification \citep{ouyang2023accounting, wang2024achieve, chen2025model}, but they have largely remained within a linear covariate-adjustment framework. However, \cite{wang2026mixed} demonstrated that covariate adjustment in such models can in fact reduce precision in certain distributions, raising questions about appropriate covariate adjustment. Moreover, because the choice of correlation structure itself can materially affect statistical efficiency, the problem of optimally learning or adapting the correlation structure remains largely unexplored. 

Recently, \cite{fang2025model} proposed a model-robust standardization framework that improves precision by leveraging parametric and semiparametric outcome models. Nevertheless, this approach does not address optimal learning of correlation structures or potential treatment-effect heterogeneity across time in stepped-wedge settings. \cite{wang2025semiparametrically, xia2025robust} derived the efficiency bounds with adjustment for cluster-level covariates only, whereas individual-level covariates contain additional contextual information that can further enhance precision but is not considered. Furthermore, none of these methods fully accommodates data-adaptive methods, such as modern machine-learning algorithms, to optimize precision gain from covariate adjustment.

In this article, we propose a new class of estimators designed to optimize precision for estimating the causal average treatment effect in stepped-wedge designs. Our approach is built on a cross-fitted partial linear outcome model that enables flexible covariate adjustment, allowing machine-learning algorithms to capture complex outcome-covariate relationships. We show that the resulting estimator is consistent and asymptotically normal provided the nuisance-function estimators converge in $L_2$ norm, even if they converge to misspecified limits. 
When the mean model is correctly specified and the nuisance estimator is consistent, the proposed estimator attains the minimal asymptotic variance under individual randomization.
To further enhance efficiency, we augment the estimator with quadratic inference functions (QIF; \citealp{qu2000improving}) to adaptively learn the correlation structure. Unlike random-effects models, which require prespecification of a single correlation structure, QIF accommodates multiple candidate structures and uses the data to optimally weight them for improved precision. In the stepped-wedge setting, we provide the first theoretical guarantee that QIF, when coupled with machine learning, yields valid asymptotic inference and never reduces precision relative to the independence working correlation. Finally, our framework accommodates treatment-effect heterogeneity across both exposure duration and calendar time, and applies seamlessly to open- and closed-cohort designs as well as the diverse design features illustrated in our motivating examples.

The remainder of this article is organized as follows. Section~\ref{sec: motivating-examples} presents two motivating examples. Section~\ref{sec: def} formally introduces the proposed causal framework. In Section~\ref{sec: estimator}, we develop the proposed estimators, and Section~\ref{sec: theory} establishes their asymptotic properties. Sections~\ref{sec: simulation} and~\ref{sec: data-application} evaluate the proposed methods through simulation studies and re-analyses of the motivating examples. Section~\ref{sec: discussion} concludes with a discussion.

\section{Motivating examples}\label{sec: motivating-examples}
\subsection{Improving caring quality for people with dementia in nursing homes using IPOS}
Patients with dementia are at high risk of receiving inadequate palliative care for their complex needs, in part because cognitive impairment often limits their ability to communicate symptoms and concerns effectively \citep{livingston2020dementia}. To address this challenge, \cite{spichiger2021improving} conducted a cluster-randomized stepped-wedge trial in Switzerland to evaluate the effectiveness of the Integrated Palliative Care Outcome Scale (IPOS), combined with subsequent case studies, in improving care for people with dementia. In this study, staff from 23 nursing homes (clusters) were equally randomized to initiate IPOS training in one of three rollout periods. The training program began at 3, 6, and 9 months after trial initiation and ultimately enrolled a total of 234 unique patients (mean cluster sizes 10.17 with standard error 3.41). We focus on the primary outcome, the QUALIDEM score measured over 3-month cycles, which is a continuous assessment of quality of care. Baseline covariates include age, sex, baseline QUALIDEM score, and baseline IPOS subscale scores for dementia physical interaction impact and dementia emotional behavioral impact.

\subsection{Procedural justice training program}
To reduce police misconduct and use of force, the Chicago Police Department implemented a procedural justice training program for its police officers \citep{wood2020procedural}. The program used an individually randomized stepped-wedge design, in which each of the 5537 qualified officers was equally randomized to one of 72 time periods (months) to receive procedural justice training. We consider the number of complaints filed against officers measured at each time period as the outcome. Baseline covariates include age, sex, baseline number of complaints, and baseline use-of-force incidents of officers.

These two data examples illustrate the stepped-wedge design across distinct research domains with markedly different sample sizes and design characteristics. Together, they highlight the methodological challenges and opportunities that arise in practice. Our goal is to develop statistical methods that improve estimation precision across such settings.
In Section~\ref{sec: data-application}, we present our re-analysis of these studies with the proposed methods.

\section{Definition and assumptions}\label{sec: def}

We consider a cluster-randomized stepped-wedge design with $I$ clusters and $J$ periods, where each cluster $i \in \{1,\dots, I\}$ contains $N_i$ individuals. The special case of individual randomization, as in the second motivating example, corresponds to $N_i \equiv 1$. The study period $j$ takes values in $\{0,1,\dots, J\}$, where $j=0$ represents the baseline period (before randomization) and $j=1,\dots, J$ represents the randomization periods. In the stepped-wedge design, each cluster is crossed over from control to treatment, and the timing of treatment onset is randomized. We use $Z_i \in \{1,\dots, J, +\infty\}$ to denote the treatment sequence, where $Z_i = j$ means the treatment starts at the beginning of period $j$. We allow for never-treated clusters, denoted as $Z_i = + \infty$.

For each individual $k$ $(k = 1,\dots, N_i)$ in cluster $i$, we denote $Y_{ijk}$ as the outcome at period $j$ and $\bX_{ik}$ as a vector of baseline variables. Here, $\bX_{ik}$ can include both individual-specific covariates and cluster-level covariates.
Since an individual may not be observed at all time points, we further define $S_{ijk}$ as the enrollment indicator at period $j$ (i.e., $S_{ijk}=1$ if $Y_{ijk}$ is observed)
and the observed cluster size  as $N_{ij} = \sum_{k=1}^{N_i} S_{ijk}$, which is no larger than the cluster-population size $N_i$. Then $N_{ij}=0$ or $1$ given individual randomization. 

Under the Neyman-Rubin \citep{neyman1990} causal inference framework, we define $Y_{ijk}(z)$ as the potential outcome had cluster $i$ been assigned to treatment sequence $z$. We assume no treatment anticipation (i.e., $Y_{ijk}(z)$ does not vary by $z$ when $z > j$) and denote $Y_{ijk}(0)$ as the potential outcome when a cluster has not been treated. Furthermore, we assume causal consistency: $Y_{ijk} = \sum_{z=1}^j I\{Z_i = z\}Y_{ijk}(z) + I\{Z_i > j\} Y_{ijk}(0)$, where $I\{\cdot\}$ is the indicator function.

Given these definitions, the observed data for each cluster are  $\bO_i=\{(Y_{ijk},Z_i,\bX_{ik}):S_{ijk}=1, j=1,\dots,J,k=1,\dots,N_i\}$, and the complete (unobserved) data are $\bW_{i}=\{(Y_{ijk}(z),S_{ijk},Z_i,\bX_{ik},N_i), k=1,\dots,N_i,0\leq z\leq j\}$. We make the following assumptions on the complete data.

\vspace{5pt}
\noindent
\textbf{Assumption 1. (Superpopulation)} (1) $\{\bW_i,i=1,..I\}$ are independent and identically distributed samples from an unknown distribution $\mathcal{P}$ with finite fourth moments. (2) Within each cluster $i$, the individual vectors $(Y_{i1k}(0),\dots,Y_{iJk}(J),\bX_{ik})$ are identically distributed given $N_i$ for $k=1,\dots,N_i$.

\vspace{5pt}
\noindent
\textbf{Assumption 2. (Randomization)}
The treatment sequence $Z_i$ is independent of all other random variables in $\bW_i$. 

\vspace{5pt}
\noindent
\textbf{Assumption 3. (Non-informative sampling)}
$\{S_{ijk} : j = 1,\dots,J,k = 1,\dots,N_i\}$ is independent of all
 other random variables in $\bW_{i}$ given $N_i$.

\vspace{5pt}

Assumption 1(1) specifies the standard sampling-based framework and the inter-cluster independence. Assumption 1(2) imposes an identical distribution on the individual-level information but allows arbitrary within-cluster correlation; this assumption is used to simplify the causal estimands. Assumption 2 is implied by the stepped-wedge design. Of note, we allow for unequal steps where $P(Z_i=j)=0$ for some $j$, which extends the classical design. Assumption 3 assumes away selection bias by imposing a random sampling scheme that is similar to missing completely at random in longitudinal data. Compared with \cite{wang2024achieve}, which imposed similar assumptions to address unequal cluster sizes across time, Assumption 3 relaxes these conditions by placing no restrictions on $N_{ij}$, while still allowing arbitrary correlation among enrollment indicators. This formulation accommodates cross-sectional, closed-cohort, and open-cohort designs \citep{li2021mixed}.

Our goal is to estimate the marginal, cluster-average treatment effect, defined as 
\begin{equation}\label{eq: estimand}
    \Delta_{j}(d)=E\left\{\frac{1}{N_i}\sum_{k=1}^{N_i}Y_{ijk}(j-d+1)\right\}-E\left\{\frac{1}{N_i}\sum_{k=1}^{N_i}Y_{ijk}(0)\right\},
\end{equation}
for $1 \le d \le j \le J$, where $d=j-z+1$ is the duration of treatment under sequence $z\le j$. It is interpreted as the effect of $d$-periods of treatment at period $j$. 
By Assumption 1(2), the estimand $\Delta_j(d)$ can be simplified to $E\{Y_{ijk}(j-d+1)\} - E\{Y_{ijk}(0)\}$.

Based on Equation~\eqref{eq: estimand}, \citet{wang2024achieve} investigated four treatment‐effect models according to whether treatment effects are homogeneous across calendar time and exposure duration.
Specifically, the \textit{constant} treatment‐effect model assumes $\Delta_j(d) \equiv \Delta$ and is widely used due to its simplicity. The \textit{duration‐specific} treatment‐effect model assumes $\Delta_j(d) \equiv \Delta(d)$ and accommodates heterogeneity with respect to exposure duration \citep{hughes2015current, kenny2022analysis, maleyeff2022assessing}, such as delayed or cumulative treatment effects. The \textit{period‐specific} treatment‐effect model assumes $\Delta_j(d) \equiv \Delta_j$, allowing treatment effects to vary over calendar time, for example, because of seasonal patterns. Finally, the \textit{saturated} treatment‐effect model imposes no structural constraints on the causal estimands and therefore involves the largest number of parameters. 
In practice, accounting for such treatment‐effect heterogeneity is critical, as ignoring it can lead to substantial bias \citep{kenny2022analysis, wang2024achieve}. Although more flexible models introduce additional parameters, interpretable summary estimands can still be reported, such as the average treatment effect across exposure durations, $\sum_{d=1}^J \Delta(d)/J$.

Table~\ref{tab:treatment-effect} summarizes the key characteristics of these models, and our proposed methods can apply to all these settings. Notably, when no units remain under control in period $J$ (as in the classical setting where $Z_i \neq +\infty$), the parameters $\Delta_J$ and $\Delta_J(d)$ are not identifiable under the period‐specific and saturated models and are therefore excluded from the set of estimands.
Finally, beyond the estimand in Equation~\eqref{eq: estimand}, \citet{fang2025model} considered constant and individual‐average treatment effects under informative cluster sizes, where individuals instead of clusters are equally weighted. Our proposed methods can be readily generalized to estimate these alternative estimands using the weighted estimating function in \citet{wang2024model}.


\begin{table}[!htbp]
    \centering
    \renewcommand{\arraystretch}{1.3}
     \begin{tabular}{
        m{3cm}    
        m{2.5cm}      
        m{4cm}      
        m{5cm}
    }
    \toprule
    \makecell[l]{Treatment effect\\ structure} & Abbreviation   &  Assumption &  \makecell[c]{Illustration}\\
\midrule
Constant     &  C & \makecell[l]{Treatment effect \\ does not vary by\\ exposure duration\\ or calendar time} & \includegraphics[width=5cm]{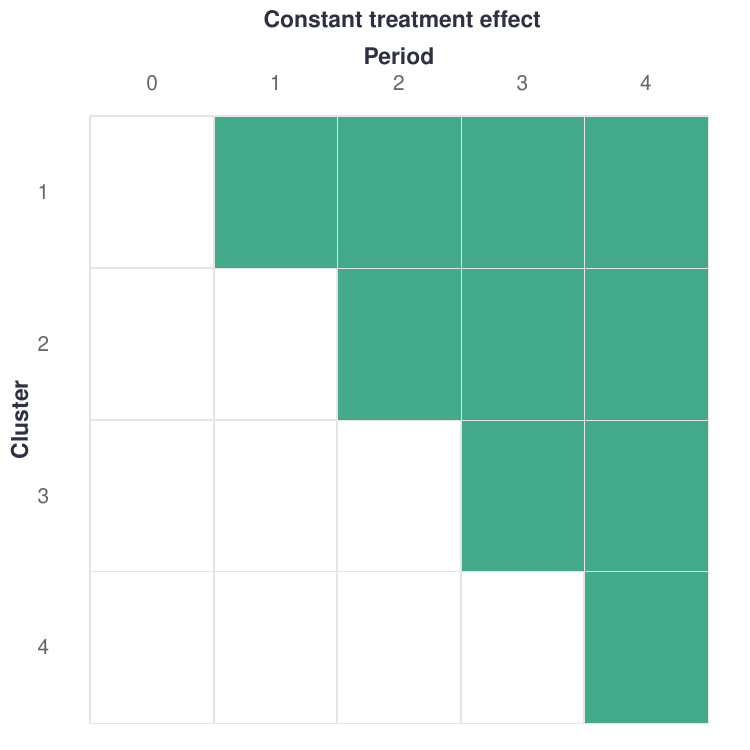}\\
    Duration-specific & D & \makecell[l]{Treatment effect\\ only varies by \\ exposure duration} & \includegraphics[width=5cm]{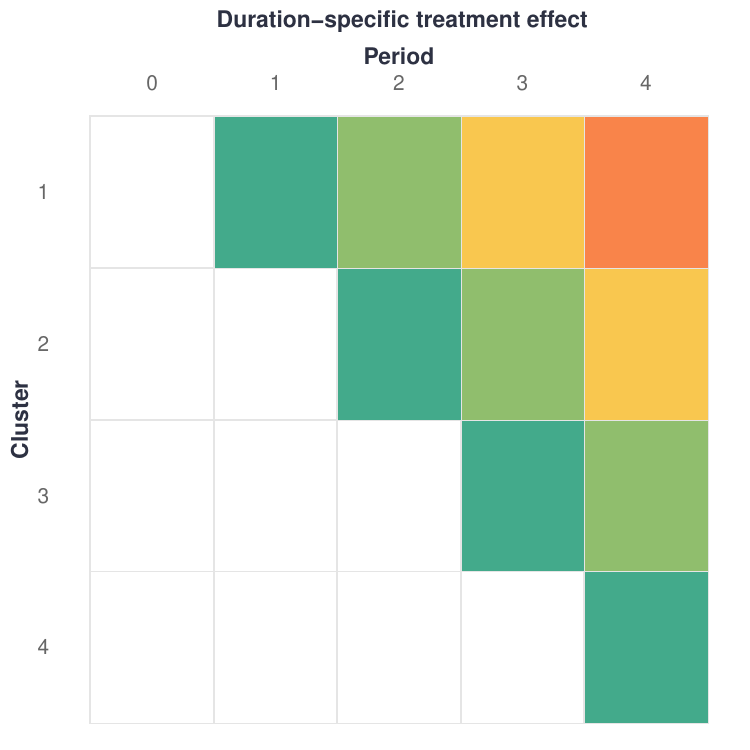}\\
    Period-specific & P & \makecell[l]{Treatment effect\\ only varies by \\ calendar time} & \includegraphics[width=5cm]{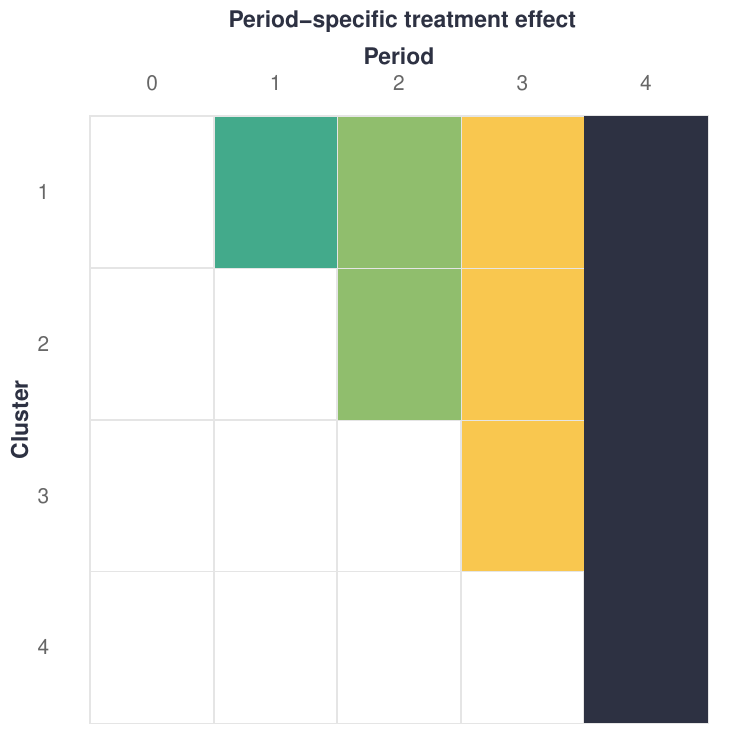}\\
    Saturated & S & \makecell[l]{Treatment effect \\ varies by \\ exposure duration\\ and calendar time} & \includegraphics[width=5cm]{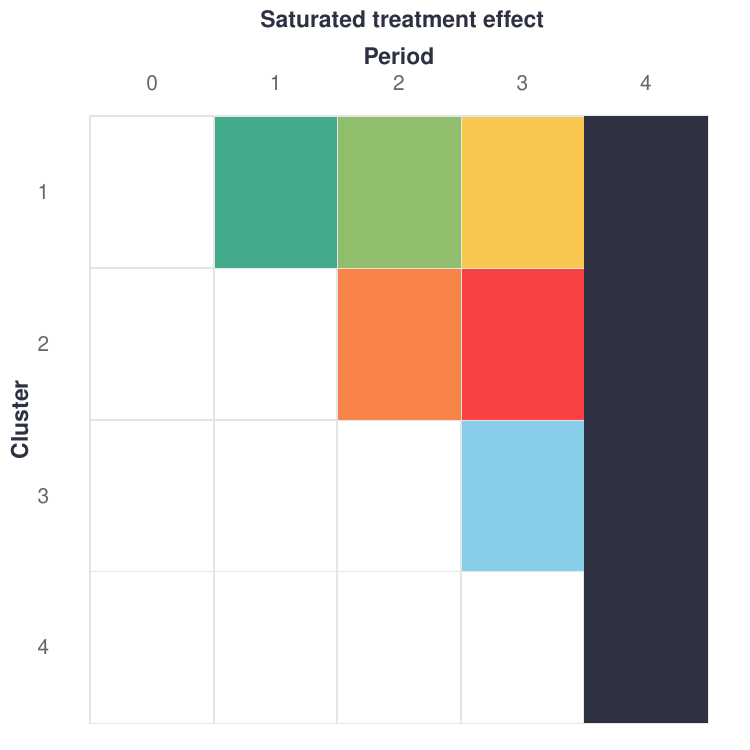}\\
    \bottomrule
    \end{tabular}
    \caption{Treatment effect structures in stepped wedge designs. In this illustration, white cells denote control conditions, whereas colored cells indicate periods under treatment. Different colors correspond to distinct treatment-effect estimands. Black cells indicate settings in which the treatment-effect estimand is not defined.}
    \label{tab:treatment-effect}
\end{table}


\section{Estimators}\label{sec: estimator}

\subsection{Flexible covariate adjustment with machine learning}
We first present the covariate adjustment method with an independence correlation structure. 
Consider a working partial linear regression model
\begin{equation}\label{eq: plr}
    E[Y_{ijk}|Z_i ,\bX_{ik}] = (\bD^*_{ij} - E[\bD^*_{ij}])\trans\bbeta^* + g_j(\bX_{ik}),
\end{equation}
where $(\bD^*_{ij} - E[\bD^*_{ij}])\trans\bbeta^*$ is the centered treatment effect term, and $g_j(\bX_{ik})=E[Y_{ijk}|\bX_{ik}]$ is the nuisance function for period $j$. In the first term, the superscript $*$ takes values in $\{\textup{C}, \textup{D}, \textup{P}, \textup{S}\}$, representing different notations when considering constant, duration-specific, period-specific, and saturated treatment effects, respectively. The notation $\bD^*_{ij}$ is a function of treatment variable $Z_{i}$ to define the appropriate treatment status, and $\bbeta^*$ is the treatment effect parameter of interest. In the second term, $g_j(\bX_{ik})$ represents the form of covariate adjustment. For example, linear adjustment corresponds to approximating $g_j(\bX_{ik})=\beta_{Xj}\trans \bX_{ik}$. 

To relate $\bbeta^*$ to our target estimands, we specific $D_{ij}^\textup{C}= I\{Z_{ij} \le j\}$, a simple indicator of receiving intervention, and $\beta^\textup{C}$ as a scalar under constant-treatment-effect model. With duration-specific treatment effect structure, we define $\bD_{ij}^\textup{D} = (I\{Z=j\}, \dots, I\{Z=1\},0,\dots,0)\trans \in \mathbb{R}^{J}$ and $\bbeta^\textup{D} = (\beta_{1}^\textup{D},\dots, \beta_{J}^\textup{D})\trans \in \mathbb{R}^{J}$. For the period-specific treatment effect, we define $D_{ij}^\textup{P}= I\{Z_{ij} \le j\}$ and use $\beta_j^\textup{P}$ as the treatment effect parameter. Finally, the saturated treatment effect structure employs $\bD_{ij}^\textup{S} = \bD_{ij}^\textup{D}$ and $\bbeta^\textup{S}_j = (\beta_{1j}^\textup{S},\dots, \beta_{J-1,j}^\textup{S})\trans \in \mathbb{R}^{J-1}$.
Together, these specifications of $\bD_{ij}^*$ and $\bbeta^*$ make sure that our estimators for $\bbeta^*$ are consistent for the estimands under each treatment effect model.

We estimate $\bbeta^*$ using a two-step cross-fitting procedure. In the first step, we apply cluster-level cross-fitting to estimate $g_j(\bX_{ik})$. Specifically, let $B_1,\dots,B_M$ denote a random partition of the $I$ clusters into $M$ folds, with each fold containing approximately $I/M$ clusters (allowing at most a difference of one in fold sizes). For each fold $m$, we use data from the remaining $M-1$ folds as the training sample to fit a prediction model for $Y_{ijk}$ given $\bX_{ik}$, yielding the fold-specific estimator $\widehat{g}_j^{(m)}$. The prediction model may take simple parametric forms, such as linear regression, or more flexible data-adaptive machine-learning methods. For observations belonging to fold $m$, the cross-fitted prediction is then given by $\widehat{g}_j^{(m)}(\bX_{ik})$.

In the second step, we treat these cross-fitted predictions as fixed and construct $\widehat{\bbeta}^*$ by minimizing the total squared loss over the observed data, i.e.,
\begin{equation}\label{eq: beta-hat}
    \widehat{\bbeta}^* = \arg \min_{\bbeta^*} \sum_{m=1}^M \sum_{i \in B_m} \sum_{j=1}^J \frac{1}{N_{ij}}\sum_{k=1}^{N_i} S_{ijk} \left\{Y_{ijk} - (\bD^*_{ij} - E[\bD^*_{ij}])\trans\bbeta^* - \widehat{g}_j^{(m)}(\bX_{ik})\right\}^2.
\end{equation}
The variance estimator for $\widehat{\bbeta}^*$, denoted as $\widehat{Var}^*(\widehat{\bbeta}^*)$, is constructed by the squared influence function for $\widehat{\bbeta}^*$, which is provided in Supplementary Material A. Of note, when $N_{ij}=0$, the term $1/N_{ij}$ is dropped from Equation~\eqref{eq: beta-hat} to avoid zeros in the denominator. 

In the above procedure, cross-fitting is employed to eliminate in-sample prediction bias when data-adaptive modeling is used for estimating $g_j(\bX_{ik})$ \citep{chernozhukov2018double}. When we adopt parametric models, such as $g_j(\bX_{ik}) = \bbeta_{Xj}\trans \bX_{ik}$, cross-fitting is unnecessary, and the resulting estimator $\widehat{\bbeta}^*$ reduces to the standard ANCOVA estimator \citep{wang2024model}. From a practical perspective, we recommend using parametric models when the effective sample size is limited (e.g., fewer than 50 observations per fold), while encouraging data-adaptive cross-fitting to optimize precision when sufficient sample sizes are available, as in our second motivating example.

In addition, the above formulation corresponds to an independence working correlation structure, since each individual $k$ and period $j$ receives equal weight. Specifically, $\widehat{\bbeta}^*$ can be equivalently characterized as the solution to
\begin{equation}\label{eq: est-eq}
    \sum_{m=1}^M \sum_{i \in B_m} (\bmD_i^* - E[\bmD_i^*\mid N_i])\trans \bmS_i\trans \bmQ_i \bmS_i
    \left\{\bY_i - (\bmD_i^* - E[\bmD_i^*\mid N_i]) \bbeta^* - \widehat{\bG}_i^{(m)} \right\} = 0,
\end{equation}
where $\bY_i$, $\bmD_i^*$, $\bmS_i$, and $\widehat{\bG}_i^{(m)}$ denote the stacked vector or matrix representations of $Y_{ijk}$, $\bmD_i^*$, $S_{ijk}$, and $\widehat{g}_j^{(m)}(\bX_{ik})$, respectively, concatenated over $j$ and $k$ (see Supplementary Material A for details). In Equation~\eqref{eq: beta-hat}, the weighting matrix $\bmQ_i$ is taken to be the identity matrix, corresponding to the independence working correlation. Alternative choices are also applicable, such as nested exchangeable correlation matrices. 



\subsection{Flexible correlation modeling with QIF}
In stepped-wedge designs, correlations may arise along multiple dimensions: across repeated visits within the same individual, among different individuals within the same cluster at a given time period, and across time within the same cluster. Because appropriately modeling the correlation structure can substantially improve statistical efficiency \citep{li2020design}, we incorporate quadratic inference functions (QIF) to adaptively learn the underlying correlation structure. Rather than prespecifying a single working correlation structure, such as the matrix $\bmQ_i$ in Equation~\eqref{eq: est-eq}, QIF represents $\bmQ_i$ as a linear combination, $\bmQ_i=\sum_{l=1}^L a_l \bmQ_{il}$, where $\bmQ_{i1}, \dots, \bmQ_{iL}$ are a collection of candidate correlation structures and $a_1,\dots, a_L$ are unknown weighting parameters estimated from the data. Each basis matrix $\bmQ_{il}$ can be chosen to capture a distinct aspect of correlation, such as an exchangeable correlation for within-cluster correlation or within-individual correlation across time, allowing QIF to adaptively combine these components to best match the observed correlation patterns.


With this adaptive correlation learning, we retain the cross-fitting procedure for covariate adjustment and modify the estimation step for $\bbeta^*$ accordingly. Specifically, we define the QIF estimator as 
\begin{equation}\label{eq: QIF}
    \widehat{\bbeta}^*_{\textup{QIF}} = \arg \min_{\bbeta^*} \left\{\sum_{i=1}^I \widetilde{\bpsi}^*_i \right\}\trans \left\{\sum_{i=1}^I \widetilde{\bpsi}^*_i(\widetilde{\bpsi}^*_i)\trans  \right\}^{-1} \left\{\sum_{i=1}^I \widetilde{\bpsi}^*_i \right\},
\end{equation}
where, for $i \in B_m$,
\begin{align*}
\widetilde{\bpsi}^*_i = \left(\begin{array}{c}
  (\bmD_i^* - E[\bmD_i^*\mid N_i])\trans \bmS_i\trans \bmQ_{i1} \bmS_i 
    \left\{\bY_i - (\bmD_i^* - E[\bmD_i^*\mid N_i]) \bbeta^* - \widehat{\bG}_i^{(m)} \right\}   \\
    \vdots \\
  (\bmD_i^* - E[\bmD_i^*\mid N_i])\trans \bmS_i\trans \bmQ_{iL} \bmS_i 
    \left\{\bY_i - (\bmD_i^* - E[\bmD_i^*\mid N_i]) \bbeta^* - \widehat{\bG}_i^{(m)} \right\}  
\end{array}\right).
\end{align*}

Compared with $\widehat{\bbeta}^*$, QIF-based estimation involves stacking estimating functions with different candidate correlation structures and solving a quadratic form of $\widetilde{\bpsi}^*$, which equivalently optimizes $(a_1,\dots, a_L)$, as shown in \cite{qu2000improving}. When a single candidate $\bmQ_{i}$ is used, then  $ \widehat{\bbeta}^*_{\textup{QIF}}$ reduces to $\widehat{\bbeta}^*$ defined in Equation~\eqref{eq: est-eq}, thereby covering the independence correlation structure as a special case. 
The variance estimator for $\widehat{\bbeta}^*_{\textup{QIF}}$, denoted as $\widehat{Var}^*(\widehat{\bbeta}^*_{\textup{QIF}})$, is obtained from the squared influence function. Full details of the variance derivation are provided in Supplementary Material A.

In practice, the choice of candidate correlation structures should be guided by the trial parameters and substantive domain knowledge. For stepped-wedge designs, commonly considered structures include independence, exchangeable correlation, and first-order autoregressive (AR-1) correlation to capture within-individual correlation across time, within-cluster within-period correlation, and within-cluster correlation across periods. We caution against unstructured correlation modeling, as it typically entails high-dimensional parameterization and substantial computational burden. Empirically, a small collection of basis matrices is often sufficient to capture most of the attainable efficiency gains \citep{qu2006quadratic}. Alternatively, data-driven correlation selection criteria similar to BIC in regression have been proposed to further enhance stability and interpretability \citep{song2009quadratic}.

\section{Asymptotic results}\label{sec: theory}
To present our theoretical statements, we denote $\bmI$ as the identity matrix and define the target the causal parameters at each treatment structure as $\Delta^C$, $\bDelta^D = (\Delta(1), \dots, \Delta(J))\trans \in \mathbb{R}^J$, $\bDelta^P = (\Delta_1, \dots, \Delta_{J-1})\trans \in \mathbb{R}^{J-1}$, and $\bDelta^S = (\Delta_1(1), \Delta_2(1), \dots, \Delta_{J-1}(J-1))\trans \in \mathbb{R}^{(J-1)J/2}$. Theorem~\ref{thm1} establishes the consistency and asymptotic normality of the proposed estimator defined in Equation~\eqref{eq: beta-hat} as the number of clusters approaches infinity.

\begin{theorem}\label{thm1}
Assuming Assumptions~1-3 and $E\left[\left\{\widehat{g}_j^{(m)}(\bX_{ik})- \underline{g}_j(\bX_{ik})\right\}^2\right] \rightarrow 0$ for some integrable limit function $\underline{g}_j(\bX_{ik})$, we have $\left\{\widehat{Var}^*(\widehat{\bbeta}^*)\right\}^{-1/2} \left(\widehat{\bbeta}^* - \bDelta^*\right) \xrightarrow{d} N(0, \bmI)$ under each treatment effect model $* \in \{C, D,P,S\}$.

If the partial linear model~\eqref{eq: plr} is correctly specified, then $\widehat{\bbeta}^* $ achieves smallest asymptotic variance if $\underline{g}_j(\bX_{ik}) = E[Y_{ijk}|\bX_{ik}]$ and $Y_{ijk}$ is independent of $\bX_{ik'}$ for $k\ne k'$ given $\bX_{ik}$.
\end{theorem}

Theorem~\ref{thm1} establishes the asymptotic validity of Wald-type inference for our proposed estimator under all treatment-effect models considered. Notably, the only requirement for nuisance-function estimation is $L_2$ convergence, and the limiting function need not be the correct target. This condition is automatically satisfied when nuisance components are estimated using parametric regression models, such as linear or spline regression. Moreover, the requirement is relatively mild for machine-learning algorithms, as it places no restriction on the rate of convergence, and is met by a wide range of methods, including random forests \citep{wager2018estimation, young2025clustered} and deep neural networks \citep{farrell2021deep}.

Theorem~\ref{thm1} further characterizes the conditions under which precision is optimized. Under individual randomization, where $k \equiv 1$, optimal precision corresponds to a correctly specified mean model and consistent estimation of the nuisance functions. The former condition is equivalent to no treatment-covariate interactions, and the latter can be achieved by many well-established machine-learning algorithms, including those mentioned above. 
For clustered data, the limit function that attains the optimal precision is $\underline{g}_j(\bX_{ik}) = E[Y_{ijk} \mid \bX_{i1}, \dots, \bX_{iN_i}]$. 
However, this target is difficult to estimate at the individual level, since all individuals within a cluster share the same covariate set. A more practical strategy is therefore to construct cluster-level summary measures and model outcomes using both individual-level covariates and these cluster-level summaries. These quantities can be incorporated into $\bX_{ik}$ in our proposed methods, yielding an estimand of the form $E[Y_{ijk} \mid \bX_{ik}]$. Accordingly, optimal precision is achieved under an additional conditional independence assumption, which yields $E[Y_{ijk} \mid \bX_{i1}, \dots, \bX_{iN_i}] = E[Y_{ijk} \mid \bX_{ik}]$ and is plausible when $\bX_{ik}$ includes sufficiently rich individual- and cluster-level covariates.

When leveraging QIF for correlation modeling, Theorem~\ref{thm2} establishes asymptotic results parallel to those in Theorem~\ref{thm1}. More importantly, it shows that QIF never decreases asymptotic precision relative to using a single working correlation structure, as is common in classical approaches. This result provides a strong theoretical guarantee that the proposed QIF method can only improve, and never harm, asymptotic precision.

\begin{theorem}\label{thm2}
Assuming Assumptions~1-3 and $ E\left[\left\{\widehat{g}_j^{(m)}(X_{ik})- \underline{g}_j(X_{ik})\right\}^2\right] \rightarrow 0$ for some integrable limit function $\underline{g}_j(X_{ik})$, we have $\left\{\widehat{Var}^*(\widehat{\bbeta}^*_{\textup{QIF}})\right\}^{-1/2} \left(\widehat{\bbeta}^*_{\textup{QIF}} - \bDelta^*\right) \xrightarrow{d} N(0, \bmI)$ under each treatment effect model $* \in \{C, D,P,S\}$.

Furthermore, $\widehat{\bbeta}^*_{\textup{QIF}}$ achieves equal or smaller variance with more basis correlation matrices. 
\end{theorem}


\begin{remark}
    In longitudinal settings, including stepped-wedge designs, semiparametric efficiency is theoretically attainable but often computationally impractical, as it requires unstructured modeling of high-dimensional covariance matrices. For example, in the absence of covariate adjustment, \citet{xia2025robust} derived the efficient score, which involves the term $Var\{\bY_i - \bmD_i^{* \top}\bbeta^*\}^{-1}$. Without imposing assumptions on the correlation structure, constructing a consistent estimator of this quantity requires empirically approximating and inverting large covariance matrices, a task that is frequently inaccurate or computationally infeasible.
    These challenges are further exacerbated in the presence of varying cluster sizes, covariate adjustment, and missing data. In practice, adopting structured working correlation models can yield satisfactory precision gains, and QIF provides an effective approach to improving correlation modeling and enhancing inferential efficiency.
\end{remark}

\section{Simulation}\label{sec: simulation}
We conduct two simulation studies to assess the finite-sample performance of the proposed estimators. The first study considers a cluster-randomized stepped-wedge trial, while the second focuses on an individually randomized stepped-wedge trial. Within each setting, we evaluate performance under both constant and duration-specific treatment-effect structures. Additional results for period-specific and fully saturated treatment-effect specifications are provided in  Supplementary Material C.

\subsection{Simulation setting}
For cluster-randomized trials, we generate 1{,}000 realizations under two trial sizes, with $I=20$ and $I=100$ clusters, representing small and moderately large trials, respectively. When $I=20$, we set the number of periods to $J=3$, the cluster-population size to $N_i=20$, and sample the observed cluster-period size $N_{ij}$ uniformly from $[5,15]$. When $I=100$, we set $J=5$, $N_i=500$, and sample $N_{ij}$ uniformly from $[5,35]$. In both scenarios, we independently generate four covariates: one cluster-level covariate $X_{i1}\sim N(0,1)$ and three individual-level covariates, $X_{ik2}\sim \mathrm{Bernoulli}(0.5)$ and $X_{ik3}, X_{ik4}\sim N(0,1)$. The sampling indicators $S_{ijk}$ follows uniform distribution given $N_{ij}$ and $N_i$. Treatment assignments $Z_i$ are generated independently across clusters, with equal probability to initiate treatment in periods $1,\dots, J$.

Under the constant treatment-effect setting, outcomes are generated according to
\begin{align}
Y_{ijk} &= I\{Z_i \le j\}\,\beta(X) + \exp(X_{ik1}X_{ik2})
+ \tfrac{1}{2}X_{ik4}^2 + I\{X_{ik4}>-1\}+ 2I\{X_{ik3}>1\}\notag \\
&\quad + I\{X_{ik1}>0.5\}(j+1) + \sigma_i + \alpha_{ij} + \tau_{ik} + \epsilon_{ijk},\label{eq: sim1}
\end{align}
where $\sigma_i,\alpha_{ij}, \tau_{ik} \sim N(0,0.1)$ denotes the cluster-level, cluster-period, individual-level random effects, respectively, and $\epsilon_{ijk} \sim N(0,0.7)$ is the independent residual error. The covariate-specific treatment effect is defined as $\beta(X) = 1 + \bigl(X_{ik3}-\overline{X}_{i\cdot 3}\bigr) + \bigl(X_{ik4}^3-\overline{X^3}_{i\cdot 4}\bigr)/2$,
where $\overline{f(X_{i\cdot p})}$ denotes the within-cluster average of $f(X_{ikp})$. Under the duration-specific treatment-effect setting, outcomes are generated as
\begin{align*}    
Y_{ijk}&=\beta_{d}(X)+\frac{1}{2}\exp{(X_{ik1}X_{ik2})}+I\{X_{ik4}>-1\}+I\{X_{ik1}>0.5\}(j+1)\\
&\quad +2I\{X_{ik3}>1\}+\sigma_i+\alpha_{ij}+\tau_{ik}+\epsilon_{ijk},
\end{align*}
where the treatment effect is
 $\beta_{d}(X)=I\{d>0\}\{1+(X_{ik3}-\overline{X_{i.3}})d/(J+1)+(X_{ik4}^3-\overline{X_{i.4}^3})/(J+1)\}$ for $d=j-Z_i+1$. 
Both outcome distributions incorporate nonlinear relationships with treatment and covariates, as well as multiple correlation structures, and are intentionally designed to demonstrate the proposed methods under challenging and realistic settings.

For individual randomized trials, we generate 1{,}000 datasets setting $I=1000$ with $J=20$, representing larger and longer trials. By design, $N_i=1$, and we set $N_{ij} \sim \textup{Bernoulli}(0.5)$ to generate unobserved outcomes in certain time points. Covariates and treatments are generated as in cluster-randomized trials. For the constant treatment effect structure, the outcome is generated as in Equation~\eqref{eq: sim1} except that cluster-level random effects are absorbed into the residual random error. For the duration-specific treatment effect structure, the outcome is generated by
\[
Y_{ijk}=\beta_d'(X)+\frac{1}{2}X_{ik3}^3+\frac{1}{2}X_{ik4}^3+\frac{J+1}{2}X_{ik1} X_{ik2}+I\{X_{ik4}>0.5\}+\tau_{ik}+\epsilon_{ijk}
\]
with $\beta_{d}'(X)=I\{d>0\}\{1+X_{ik3}-\overline{X_{i.3}}+(X_{ik4}^3-\overline{X_{i.4}^3})d\}$ for $d=j-Z_i+1$.

For each simulated dataset, we apply our proposed approach to estimate the treatment effect estimands. Under a working independence working correlation, we implement the estimator defined in Equation~\eqref{eq: beta-hat} with three covariate adjustment options: no adjustment, linear adjustment, and machine-learning-based adjustment. Linear adjustment specifies $\widehat{g}_j(\bX_{ik}) = \bbeta_{Xj}\trans \bX_{ik}$, and machine-learning adjustment uses an ensemble of regression trees \citep{breiman2017classification}, random forests \citep{breiman2001random}, generalized linear models, and support vector machines \citep{cortes1995support}, implemented with the \texttt{SuperLearner} R package \citep{van2007super}. We further implement QIF-assisted estimators, for which the covariate adjustment strategies remain the same, while the basis correlation structures are chosen to reflect the underlying random-effects components.
When $I=20$, we apply the 4MD correction of \citep{westgate2012bias} to improve variance estimation for QIF in small samples. To benchmark the proposed methods, we also fit linear mixed-effects models with correctly specified random-effects structures and linear adjustment for baseline covariates; model-based variance estimator from the R function \texttt{lme4} \citep{bates2015package} was used for convenience. For each estimator, we report the bias, empirical standard error, average estimated standard error, and the coverage probability of 95\% confidence intervals constructed using a $t$-distribution with $I-p$ degrees of freedom, where $p$ denotes the dimension of the treatment effect vector.

\subsection{Simulation Results}
Table~\ref{tab:1} summarizes the simulation results under the constant treatment-effect setting for cluster-randomized trials with $I=20,100$ clusters and individually randomized trials with $I=1000$ individuals. Across all scenarios, the proposed estimators exhibit negligible bias and achieve close-to-nominal 95\% coverage, supporting our asymptotic theory and demonstrating satisfactory finite-sample performance, even when the number of clusters is small. Relative to the unadjusted estimator, linear covariate adjustment reduces variance by 25-70\%, and machine-learning-based adjustment consistently achieves further precision gain, particularly in larger samples (over 90\% variance reduction) where sufficient data mitigate overfitting. These patterns are consistent with the fact that the benefits of flexible learning become more pronounced as the sample size increases. Incorporating QIF to adaptively learn the correlation structure yields additional efficiency improvements across all settings, with variance reductions of roughly 50-70\% compared with the independence working correlation. Taken together, these results highlight the complementary roles of flexible covariate adjustment and adaptive correlation learning in improving efficiency.

The benchmark linear mixed-effects model performs comparably to the proposed estimator with QIF and linear adjustment, indicating that our framework is compatible with simple parametric adjustment. However, in the large-sample setting ($I=1000$), the linear mixed-effects model exhibits noticeable undercoverage (0.812), reflecting bias in the model-based variance estimator under model misspecification; this issue can be alleviated by using model-robust variance estimators \citep{wang2024achieve}. Moreover, while our implementation of the mixed-effects model assumes correct specification of the random-effects structure, such assumptions are often violated in practice, potentially leading to efficiency loss. In contrast, QIF flexibly combines multiple candidate correlation structures, increasing the chance of capturing the true dependence structure and yielding more robust efficiency gains.



For the duration-specific treatment-effect setting, there are $J$ causal estimands corresponding to different exposure durations. For brevity, we report only the average duration-specific effect, defined as $\sum_{d=1}^{J}\Delta(d)/J$. Table~\ref{tab:2} summarizes the simulation results, which further demonstrate the validity of the proposed methods in this more complex setting and show comparable precision gains from machine-learning-based covariate adjustment and adaptive correlation modeling to those observed in Table~\ref{tab:1}.


In Supplementary Material~C, we present parallel results under the period-specific and saturated treatment-effect structures, which yield similar findings. To further assess the performance of individual machine-learning methods, we repeat the simulation studies using 8 standalone algorithms. The results, reported in Supplementary Material C, show that regression trees, random forests, and support vector machines achieve the best overall performance under the data-generating mechanisms considered. These algorithms are therefore included in the ensemble learner used in our analyses.



 \begin{table}[t]
		\centering  
   \caption{Summary results for estimating the constant treatment effect. For cluster-randomized trials with $I=20,100$ clusters and individual-randomized trials with $I=1000$ individuals, we report the bias, empirical standard error (ESE), average of estimated standard errors (ASE), and coverage probability of the 95\% confidence interval (CP).  }
		\setlength{\tabcolsep}{5mm}{
        \renewcommand{\arraystretch}{1.3}
			\begin{tabular}{ccccccc}
				\toprule
				 $I$  & \makecell{Correlation \\ modeling} & \makecell{Covariate \\ adjustment} &  Bias       &  ESE  &  ASE   &  CP    
				\\ \midrule
				\multirow{7}{*}{20} & & Unadjusted&-0.051&0.872&0.859&0.940\\
                & Independence&Linear & -0.011 & 0.614     & 0.547      & 0.942 \\
                    &&ML&-0.024 & 0.474 &0.415  &0.952 \\
                    \cline{2-7}
                 &&Unadjusted&0.021&0.313&0.327&0.954\\
                &QIF& Linear &0.017 & 0.271     & 0.263
                &  0.952 \\
                    &&ML& 0.002 &0.248&0.225  &0.936  \\
                    \cline{2-7}
                   &Random effects &Linear& 0.018  &0.252  &0.263  & 0.970  
				\\ \midrule
                \multirow{7}{*}{100} &&Unadjusted&-0.016&0.436&0.431&0.956\\
                & Independence &Linear& 0.002 & 0.242      &  0.241      & 0.952  \\
                    &&ML& -0.003 & 0.123 & 0.112 & 0.959\\ \cline{2-7}
                 &&Unadjusted&0.014&0.136&0.140&0.958\\
                &QIF & Linear&0.007 & 0.094     & 0.095
                &  0.950 \\
                    &&ML& 0.006 &0.057& 0.057 & 0.944  \\ \cline{2-7}
                    &Random effects&Linear&0.007&0.124&0.129&0.963
				\\ \midrule
                \multirow{7}{*}{1000}&&Unadjusted&0.006&0.347&0.353&0.952\\
                & Independence &Linear& 0.005 & 0.209     & 0.219      & 0.965 \\
                    &&ML& 0.000 &0.063  & 0.058 &0.961 \\ \cline{2-7}
                 &&Unadjusted&0.014&0.181&0.177&0.946\\
                & QIF &Linear&0.003 & 0.109     & 0.112                & 0.952  \\
                    &&ML&0.000 &0.037& 0.034 &0.934  \\
                    \cline{2-7}
                    &Random effects&Linear&0.010 &0.138  & 0.095 & 0.812
				\\ \bottomrule
		\end{tabular}}
        \label{tab:1}
	\end{table}

   \begin{table}[!htbp]
		\centering
\caption{Summary results for estimating the averaged duration-specific treatment effect. For cluster-randomized trials with $I=20,100$ clusters and individual-randomized trials with $I=1000$ individuals, we report the bias, empirical standard error (ESE), average of estimated standard errors (ASE), and coverage probability of the 95\% confidence interval (CP). }
\setlength{\tabcolsep}{5mm}{
\renewcommand{\arraystretch}{1.3}
\begin{tabular}{ccccccc}
\toprule
 $I$  & \makecell{Correlation \\ modeling} & \makecell{Covariate \\ adjustment} & Bias & ESE & ASE & CP \\
\midrule
\multirow{7}{*}{20} &  & Unadjusted & -0.037 & 1.081 & 1.085 & 0.934 \\
   &  Independence & Linear     &  0.005 & 0.712 & 0.700 & 0.947 \\
   &  & ML         & -0.006 & 0.518 & 0.462 & 0.957 \\
   \cline{2-7}
   &          & Unadjusted & -0.002 & 0.767 & 0.803 & 0.943 \\
   & QIF          & Linear     & -0.003 & 0.581 & 0.572 & 0.949 \\
   &           & ML         &  0.010 & 0.377 & 0.328 & 0.943 \\
   \cline{2-7}
   & Random effects         & Linear     & -0.007 & 0.552 & 0.459 & 0.911 \\
\midrule
\multirow{7}{*}{100} & & Unadjusted &  0.032 & 0.595 & 0.611 & 0.953 \\
    & Independence & Linear     &  0.021 & 0.361 & 0.360 & 0.942 \\
    &  & ML         &  0.004 & 0.143 & 0.124 & 0.960 \\
    \cline{2-7}
    &           & Unadjusted  & -0.034 & 0.381 & 0.351 & 0.928 \\
    & QIF          & Linear    &  0.003 & 0.253 & 0.241 & 0.944 \\
    &           & ML         &  0.006 & 0.071 & 0.069 & 0.939 \\
    \cline{2-7}
    & Random effects         & Linear     & -0.015 & 0.309 & 0.213 & 0.824 \\
\midrule
\multirow{7}{*}{1000} &  & Unadjusted & -0.014 & 0.479 & 0.484 & 0.945 \\
     & Independence & Linear     & -0.008 & 0.327 & 0.327 & 0.949 \\
     &  & ML         & -0.003 & 0.138 & 0.140 & 0.953 \\
     \cline{2-7}
     &           & Unadjusted & -0.001 & 0.399 & 0.394 & 0.942 \\
     & QIF          & Linear     & -0.007 & 0.276 & 0.279 & 0.946 \\
     &           & ML         & -0.004 & 0.077 & 0.075 & 0.954 \\
     \cline{2-7}
     & Random effects        & Linear     &  0.044 & 0.295 & 0.116 & 0.558 \\
\bottomrule
\end{tabular}}

        \label{tab:2}
	\end{table} 

\section{Data application}\label{sec: data-application}

\begin{figure}[t]  
    \centering
    \includegraphics[width=\textwidth]{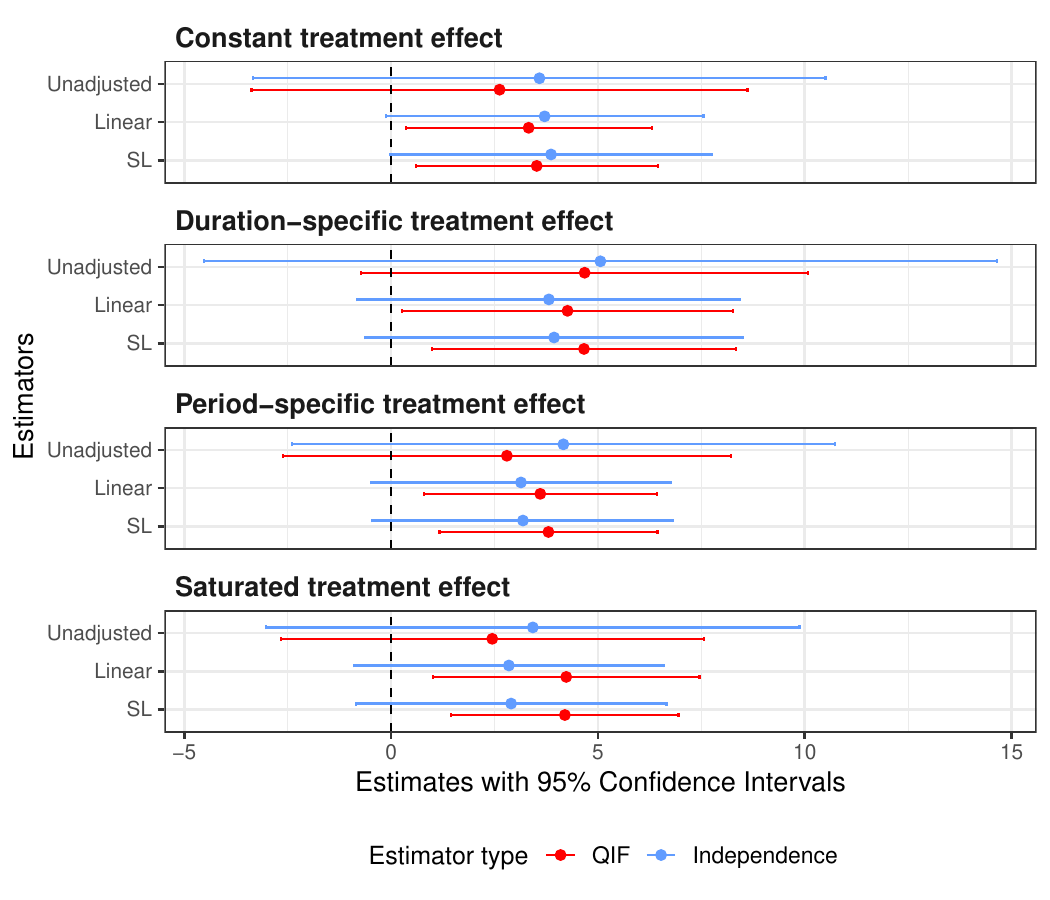} 
    \caption{Point Estimates with 95\% confidence intervals of IPOS data. For duration-specific, period-specific, and saturated treatment structures, results for the averaged treatment effects across estimands are presented.}
    \label{fig:1}
\end{figure}

\begin{figure}[t]  
    \centering
    \includegraphics[width=\textwidth]{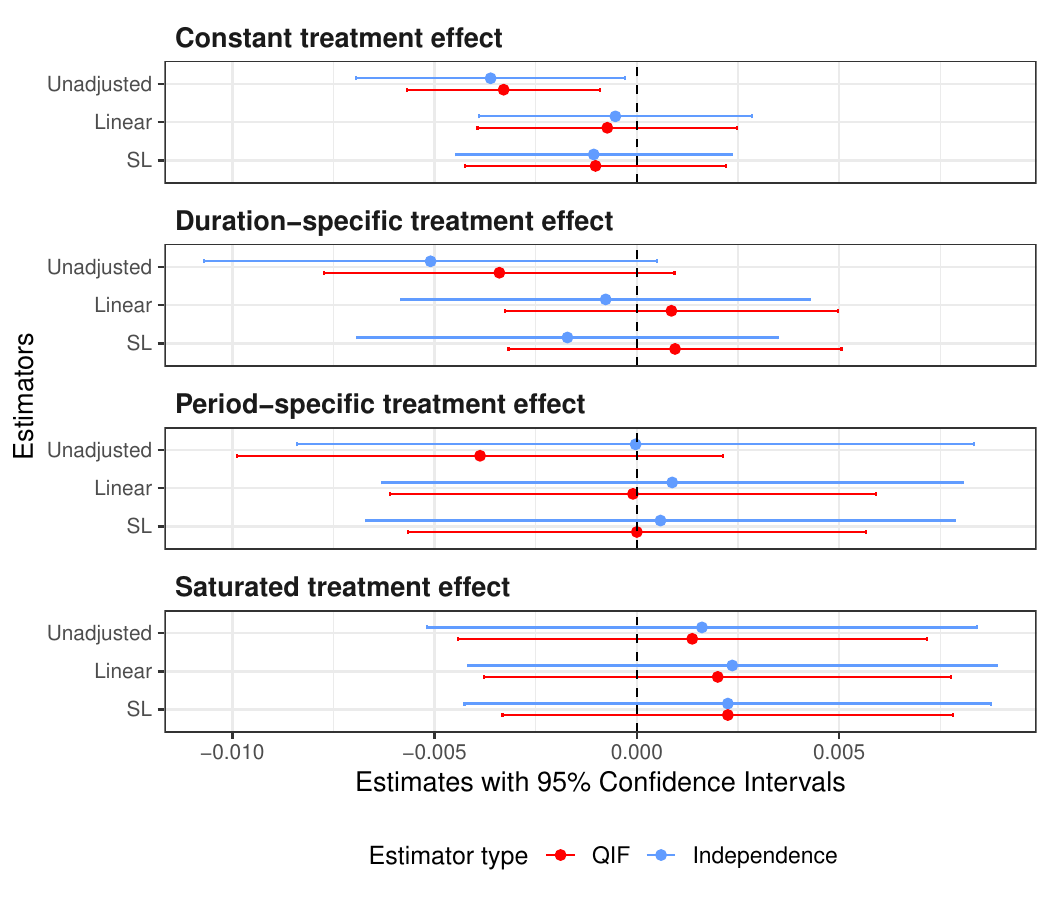} 
    \caption{Point estimates with 95\% confidence intervals of Chicago Police Training Data. For duration-specific, period-specific, and saturated treatment structures, results for the averaged treatment effects across estimands are presented.}
    \label{fig:2}
\end{figure}

We apply the proposed methods to the two motivating examples using the same specifications as in the simulation studies. We do not implement mixed-effects models because their model-based inference can lead to undercoverage, as demonstrated in our simulation results. When treatment effects vary across exposure duration or calendar time, we report averaged treatment effects for brevity. Figures~\ref{fig:1} and~\ref{fig:2} display the point estimates and 95\% confidence intervals for each estimator in the first and second motivating examples, respectively.


Figure~\ref{fig:1} shows beneficial effects of IPOS in improving patient care across all specifications. Both QIF and covariate adjustment yield substantial variance reductions (28-70\%), resulting in statistically significant treatment effects at the 5\% level, whereas the unadjusted analyses fail to reach significance. These findings further highlight the practical value of the proposed methods. Moreover, despite the limited sample size ($I=20$), machine-learning-based covariate adjustment achieves an additional 0-18\% variance reduction relative to linear adjustment across settings, illustrating the finite-sample benefits of data-adaptive estimation in stepped-wedge designs.

In Figure~\ref{fig:2}, the unadjusted analysis suggests that procedural justice training reduces complaints, whereas this apparent effect disappears after covariate adjustment. This pattern is consistent with findings reported in \cite{wood2021revised}. Our further investigation indicates that the nominal significance in the unadjusted analysis is driven by imbalance in baseline covariates. Specifically, officers assigned to training exhibit lower baseline use-of-force than those in the control group (See Supplementary Material D for a visualization). Because this baseline variable is prognostic of the outcome, such imbalances induce an estimate shift in the treatment effect estimate, which would be eliminated under a balanced design by covariate adjustment \citep{wang2019analysis}.
In terms of efficiency, QIF improves precision by 20-45\%, whereas covariate adjustment alone yields only moderate gains of 8-25\%. Moreover, machine learning offers little additional variance reduction beyond linear covariate adjustment 0-6\%, suggesting that the outcome–covariate relationship is not strongly nonlinear in this setting.

\section{Discussion}\label{sec: discussion}

Motivated by the increasing use of stepped-wedge designs across diverse applied domains with heterogeneous design features, we develop flexible, efficient, and robust methods for causal inference under multiple treatment-effect structures. Our theoretical results demonstrate that integrating modern machine-learning-based covariate adjustment with adaptive correlation learning yields asymptotic efficiency gains while preserving valid inference. Simulation studies and real-data applications further demonstrate the finite-sample robustness and practical advantages of the proposed approach. Together, these features offer a principled and practically implementable alternative to standard mixed-effects and parametric regression analyses for stepped-wedge designs, particularly in settings where correlation structures are complex, covariate-outcome relationships are nonlinear, and treatment effects vary over exposure duration and calendar time.

Our framework is developed under simple randomization, whereas covariate-adaptive randomization schemes, such as stratified randomization \citep{Zelen1974} and covariate-constrained randomization \citep{morgan2012rerandomization}, are commonly used to improve baseline balance and enhance precision. Building on the theoretical results of \cite{wang2023model, wang2024asymptotic}, the proposed methods extend directly to these settings without compromising the asymptotic validity. In particular, machine-learning-based covariate adjustment combined with standard cross-fitting remains asymptotically valid and can fully capture the efficiency gains induced by covariate-adaptive randomization, provided that the covariates used in the randomization scheme are included in the adjustment set.

To accommodate missing outcomes and time-varying cluster sizes in stepped-wedge designs, we adopt a noninformative sampling assumption to target the population-level causal estimands. An alternative approach is to target the observed population by conducting a cluster-level analysis based on within-cluster averaged outcomes, for which the proposed methods remain directly applicable. When outcome missingness depends on observed covariates, the missingness mechanism can be modeled to construct doubly robust estimators within a longitudinal data framework \citep{tsiatis2007}. A formal development and evaluation of such extensions in stepped-wedge settings is an important direction for future research.

\section*{Acknowledgements}


Research reported in this publication was supported by the National Institute Of Allergy And Infectious Diseases of the National Institutes of Health under Award Number R00AI173395. The content is solely the responsibility of the
authors and does not necessarily represent the official views of the National Institutes of Health.\vspace*{-8pt}


\section*{Supplementary Materials}

Web Appendices, Tables, and Figures referenced in Sections 3-6 and code are available with this paper at the Biometrics website on Oxford Academic.\vspace*{-8pt}



\bibliographystyle{apalike}
\bibliography{library}

\end{document}


\title{\Huge Supplementary Material to ``Optimizing precision in stepped-wedge designs via machine learning and quadratic inference function''}   
\author{}
\date{}

\maketitle

Supplementary Material A provides the estimating equations and variance estimators. Supplementary Material B provides the proof for our theorems. Supplementary Material C provides the additional simulation results. Supplementary Material D visualizes the baseline imbalance of the Chicago training program. 

\section{Estimating equations and variance estimators}
We first present the estimating equations for computing $\widehat{\bbeta}^*$.

We define $\bY_i=(Y_{i11},Y_{i12},\dots,Y_{iJN_i})\trans\in \RR^{J N_i}$ and $\bG_i=(g_1(\bX_{i1}),g_1(\bX_{i2}),\dots,g_{J}(\bX_{iN_i}))\trans \in \RR^{JN_i}$. In addition, let $\bmS_{ij}\in \RR^{N_{ij}\times N_i}$ represents the missing matrix for cluster $i$ at time $j$. Specifically, $\bmS_{ij}$ is constructed by extracting all $k$-th rows  of the identity matrix $\bmI_{N_i}$ corresponding to $k: S_{ijk}=1$. Let $\bmS_i=bdiag(\bmS_{i1}/N_{i1}^{1/2},\dots,\bmS_{iJ}/N_{iJ}^{1/2})\in \RR^{(\sum N_{ij})\times N_iJ}$, which implies that $\bmS_i$ is the function of $N_i$ and $S_{ijk}$. We next define $\bmD_i^*$ as the stacked matrix of $\bD_{ij}^*$ across individuals. Under constant treatment effect structure, we define $\bmD_i^C=\bH_i^C\otimes\bone_{N_i}\in \RR^{N_iJ}$ for $\bH_i^C=(I\{Z_i\leq 1\},\dots,I\{Z_i\leq J\})\trans$, where$\bone_{N_i}$ denotes a $N_i$-dimensional vector of ones and `$\otimes$' is the Kronecker product. Under duration-specific treatment effect structure, we define $\bmD_i^D=\bmH_i^D\otimes\bone_{N_i}\in \RR^{JN_i\times J}$ where 
\[
\bmH_i^D = 
\left(
\begin{array}{cccc}
I\{Z_i = 1\} & & & \\
I\{Z_i = 2\} & I\{Z_i = 1\} & & \\
\vdots & \vdots & \ddots & \\
I\{Z_i = J\} & I\{Z_i = J-1\} & \cdots & I\{Z_i = 1\}
\end{array}
\right).
\]
Under period-specific treatment effect structure, we denote $\bmD_i^P=\bmH_i^P\otimes\bone_{N_i}\in\RR^{(J-1)N_i\times(J-1)}$ for $\bmH_i^D=diag(I\{Z_i\leq 1\},\dots,I\{Z_i\leq (J-1)\})$. Under the saturated treatment effect strucutre, we define $\bmD_i^S=\bmH_i^S\otimes\bone_{N_i}\in \RR^{N_I\{J-1\}\times (J-1)J/2}$, where $\bmH_i^S=diag(\bH_{i1}\trans,\dots,\bH_{i,J-1}\trans)$ and $\bH_{ij}=(I\{Z_i=j\},\dots,I\{Z_i=1\})\trans$. 

With the above notation, the estimating function is defined as
\[
 \bpsi_i^*(\bbeta^*,\bG_i)=(\bmD_i^*-\bmu_i^*)\trans\bmS_i\trans \bmS_i\{\bY_i-(\bmD_i^*-\bmu_i^*\}\bbeta^*-\bG_i),    
\]
where $\bmu_i^*=E(\bmD_i^*\mid N_i)$. Here, $\bmu_i^*$ is conditioned on $N_i$ because its dimension varies by $N_i$; at the individual level, $N_i$ is independent of $\bD_{ij}^*$ due to randomization.
The proposed $\widehat{\bbeta}^*$ is then computed from the estimating equation $\sum_{m=1}^{M}\sum_{i\in B_m}  \bpsi_i^*(\bbeta^*,\widehat\bG^{(m)}_i) = \bzero$, where $\widehat\bG^{(m)}_i= (\widehat g_{1}^{(m)}(\bX_{i1}),\widehat g_{1}^{(m)}(\bX_{i2}),\dots, \widehat g_{J}^{(m)}(\bX_{iN_i}))\trans \in \mathbb{R}^{JN_i}$ for $i \in B_m$.

The variance estimator, $\widehat{Var}^*(\widehat{\bbeta}^*)$, is defined as
\begin{equation}\label{var-est: 1}
\widehat{\bSigma}^*=\left(\widehat\bmV^*\right)^{-1}\left\{\frac{1}{M}\sum_{m=1}^{M}\frac{1}{n}\sum_{i\in B_m}\bpsi_i^*(\widehat{\bbeta}^*,\widehat\bG^{(m)}_i)\bpsi_i^*(\widehat{\bbeta}^*,\widehat\bG^{(m)}_i)\trans\right\}\left(\widehat\bmV^*\right)^{-1},
\end{equation}
where 
\[
\widehat\bmV^*=\frac{1}{I}\sum_{i=1}^{I}(\bmS_i\bmD_i^*-\bmS_i\bmu_i^*)\trans(\bmS_i\bmD_i^*-\bmS_i\bmu_i^*).
\]

Next, for QIF-based estimation, we define 
\begin{align*}
\bpsi_I^*(\bbeta^*,\widehat{\bG}) = \frac{1}{I}\sum_{i=1}^I\widetilde{\bpsi}^*_I\{\bbeta^*,\widehat{\bG}_i^{(m\}}) = \frac{1}{I}\sum_{i=1}^I\left(\begin{array}{c}
  (\bmD_i^* -\bmu^* )\trans \bmS_i\trans \bmQ_{i1} \bmS_i 
    \left\{\bY_i - (\bmD_i^* - \bmu^*) \bbeta^* - \widehat{\bG}_i^{(m)} \right\}   \\
    \vdots \\
  (\bmD_i^* - \bmu^*)\trans \bmS_i\trans \bmQ_{iL} \bmS_i 
    \left\{\bY_i - (\bmD_i^* - \bmu^*) \bbeta^* - \widehat{\bG}_i^{(m)} \right\}  
\end{array}\right).
\end{align*}
Then, the estimator in the main paper can be reformulated as
\[
\widehat{\bbeta}^*_{\textup{QIF}} = \arg \min_{\bbeta^*}  \bpsi_I^*(\bbeta^*,\widehat{\bG})\trans \left\{\bmC_I^*(\bbeta^*,\widehat{\bG}) \right\}^{-1} \bpsi_I^*(\bbeta^*,\widehat{\bG}),
\]
where $\bmC_I^*(\bbeta^*,\widehat{\bG})=I^{-1}\sum_{i=1}^{I}\widetilde{\bpsi}^*_I\{\bbeta^*,\widehat{\bG}\}\widetilde{\bpsi}^*_I\{\bbeta^*,\widehat{\bG}\}\trans$. 
The variance estimator, $\widehat{Var}^*(\widehat{\bbeta}^*_{\textup{QIF}})$, is defined as
\begin{equation}\label{var-est: QIF}
\widehat\bSigma^*_{\textup{QIF}}=\left(\frac{\partial \bpsi^*_I\{\widehat{\bbeta}^*_{\textup{QIF}},\widehat\bG\}}{\partial (\bbeta^*)\trans}\{\bmC^*_I\{\widehat{\bbeta}^*_{\textup{QIF}},\widehat\bG\}\}^{-1}\frac{\partial \bpsi^*_I\{\widehat{\bbeta}^*_{\textup{QIF}},\widehat\bG\}}{\partial \bbeta^*}\right)^{-1},
\end{equation}
where
\[
\frac{\partial \bpsi^*_I\{\widehat{\bbeta}^*_{\textup{QIF}},\widehat\bG\}}{\partial \bbeta^*}=-\frac{1}{I} \sum_{i=1}^I
\begin{pmatrix}
 (\bmD_i^*-\bmu^*)\trans \bmS_i\trans\bmQ_{i1}\bmS_i\{\bmD_i^*-\bmu^*\}\\
\vdots \\
 (\bmD_i^*-\bmu^*)\trans \bmS_i\trans\bmQ_{iL}\bmS_i\{\bmD_i^*-\bmu^*\}
\end{pmatrix}.
\]

\section{Proofs}
\subsection{Proof for Theorem 1}

We inherit all notation from Section A. In this proof, we first establish that solving function $E(\bpsi_i^*(\bbeta^*,\bG_i))=\bzero$ yields $\bbeta^*=\bDelta^*$.
We next show that, for any integrable limiting function $\underline\bG_i= (\underline{g}_1(\bX_{i1}),\underline{g}_1(\bX_{i2}),\dots, \underline{g}_J(\bX_{iN_i}))\trans \in \mathbb{R}^{JN_i}$, we have $E( \bpsi_i^*(\bDelta^*,\underline\bG_i))=\bzero$.  Then, we prove the consistency and asymptotic normality of $\widehat{\bbeta}^*$. Next, we established the consistency of our variance estimator defined in Equation~\eqref{var-est: 1}. Finally, we characterize when the estimator achieves the highest precision. Throughout the proof, $\| \cdot\|$ denotes the Euclidean norm for vectors and the $L_2$ norm for matrices.

\vspace{0.2in}
\noindent\textbf{Part I (Unbiased estimating equation)}


Under the constant treatment effect structure, we define the conditional treatment effects function for individual $k$ in cluster $i$ as
\[
h_j^C(\bX_{ik})=E[Y_{ijk}\mid \bX_{ik},D^C_{ij}=1]-E[Y_{ijk}\mid \bX_{ik},D^C_{ij}=0]
\]
and define$\mu^C_j=ED^C_{ij}$. Then, denoting
\[
\epsilon_{ijk}=Y_{ijk}-(D^C_{ij}-\mu^C_j)h_j^C(\bX_{ik})-g_j(\bX_{ik}),
\]
Assumption 2 yields 
\begin{align*}
E[\epsilon_{ijk}\mid\bX_{ik},D_{ij}^C] &= D_{ij}^C E[Y_{ijk}\mid \bX_{ik},D^C_{ij}=1] + (1-D_{ij}^C )E[Y_{ijk}\mid \bX_{ik},D^C_{ij}=0] \\
&\quad - (D^C_{ij}-\mu^C_j)h_j^C(\bX_{ik})-g_j(\bX_{ik}) \\
&= \mu^C_j E[Y_{ijk}\mid \bX_{ik},D^C_{ij}=1] + (1-\mu^C_j) E[Y_{ijk}\mid \bX_{ik},D^C_{ij}=0] - E[Y_{ijk}\mid \bX_{ik}] \\
&=0.
\end{align*}
We can rewrite the estimating equation as
\begin{align*}
    &E\bpsi_i^C(\bbeta^C,\bG_i)=\sum_{j=1}^{J}E \left[N_{ij}^{-1}\sum_{k=1}^{N_i}S_{ijk}(D^C_{ij}-\mu^C_j)\{Y_{ijk}-(D^C_{ij}-\mu^C_j)\beta^C-g_j(\bX_{ik})\}\right]\\
    &=\sum_{j=1}^{J}E\left[N_{ij}^{-1}\sum_{k=1}^{N_i}S_{ijk}(D^C_{ij}-\mu^C_j)\{\epsilon_{ijk}+(D^C_{ij}-\mu^C_j) (h_j^C(\bX_{ik})-\beta^C)\}\right]\\
    &=\sum_{j=1}^{J}E\left[\sum_{k=1}^{N_i}E[N_{ij}^{-1}S_{ijk}|N_i]E[(D^C_{ij}-\mu^C_j)\{\epsilon_{ijk}+(D^C_{ij}-\mu^C_j) (h_j^C(\bX_{ik})-\beta^C)\}|N_i]\right]\\
    &= \sum_{j=1}^{J}E\left[(D^C_{ij}-\mu^C_j)\{\epsilon_{ijk}+(D^C_{ij}-\mu^C_j) (h_j^C(\bX_{ik})-\beta^C)\right]\\
    &=\sum_{j=1}^{J}\Var(D^C_{ij})E[\{h_j^C(\bX_{ik})-\beta^C\}],
\end{align*}
where the second line comes from the definition of $\epsilon_{ijk}$, the third line results from non-informative sampling (Assumption 3), the fourth line is implied by Assumption 1(2) and $\sum_{k=1}^{N_i}E[N_{ij}^{-1}S_{ijk}|N_i] = 1$, and the last line uses $E[\epsilon_{ijk}\mid\bX_{ik},D_{ij}^C] = 0$ and randomization (Assumption 2).

By setting $E\bpsi_i^C(\bbeta^C,\bG_i)=0$ and denoting $w_j = \Var(D^C_{ij})/(\sum_{j=1}^J \Var(D^C_{ij}))>0$, we have that
\begin{align*}
\beta^C&=\frac{1}{J}\sum_{j=1}^J w_j E[h_j^C(\bX_{ik})]\\
&=\frac{1}{J}\sum_{j=1}^Jw_j \left (E[Y_{ijk}\mid D^C_{ij}=1]-E[Y_{ijk}\mid D^C_{ij}=0]\right)\\
&= \frac{1}{J}\sum_{j=1}^Jw_j \left (\sum_{z=1}^j E[I\{Z_i=z\}Y_{ijk}(z)\mid Z_i \le j] - E[I\{Z_i>j\} Y_{ijk}(0)\mid Z_i > j]\right) \\
&= \frac{1}{J}\sum_{j=1}^Jw_j\left(\sum_{z=1}^j P(Z_i=z|Z_i\le j) E[Y_{ijk}(z)] - E[Y_{ijk}(0)]\right)\\
&= \frac{1}{J}\sum_{j=1}^Jw_j\sum_{z=1}^j P(Z_i=z|Z_i\le j) \Delta_j(j-z+1)\\
&=\Delta^C,
\end{align*}
where the last line uses the constant treatment effect structure that $\Delta_j(j-z+1) \equiv \Delta$.

Similarly, under the duration-specific treatment effect model, we define $\bh_j^D(\bX_{ik})=(h_j^D(\bX_{ik};d=1),\dots,h_j^D(\bX_{ik};d=j),0,\dots, 0)\trans \in \mathbb{R}^J$ where for $d=1,\dots,j$,
\[
h_j^D(\bX_{ik};d)=E[Y_{ijk}\mid \bX_{ik},Z_i=j-d+1]-E[Y_{ijk}\mid \bX_{ik},Z_i> j].
\]
Then, for $\bmu_j^D=E(\bD_{ij}^D)$ and
\[
\epsilon_{ijk}=Y_{ijk}-(\bD^D_{ij}-\bmu^D_j)\trans \bh^D_j(\bX_{ik})-g_j(\bX_{ik}),
\]
we have 
\begin{align*}
    & E[\epsilon_{ijk}\mid \bX_{ik},\bD_{ij}^D]\\
    &= \sum_{d=1}^j I\{Z_i=j-d+1\} E[Y_{ijk}\mid \bX_{ik},Z_i=j-d+1] + I\{Z_i> j\}E[Y_{ijk}\mid \bX_{ik},Z_i>j] \\
    &\quad -(\bD^D_{ij}-\bmu^D_j)\trans \bh^D(\bX_{ik})-g_j(\bX_{ik})\\
    &= \sum_{d=1}^j P(Z=j-d+1) E[Y_{ijk}\mid \bX_{ik},Z_i=j-d+1] + P(Z>j) E[Y_{ijk}\mid \bX_{ik},Z_i>j] - E[Y_{ijk}|\bX_{ik}]\\
    &=0.
\end{align*}
Then, we derive 
\begin{align*}
    &E\bpsi_i^D(\bbeta^D,\bG_i)= \sum_{j=1}^{J}E \left[N_{ij}^{-1}\sum_{k=1}^{N_i}S_{ijk}(\bD^D_{ij}-\bmu^D_j)\{Y_{ijk}-(\bD^D_{ij}-\bmu^D_j)\trans\bbeta^D-g_j(\bX_{ik})\}\right]\\
    &=\sum_{j=1}^{J}E\left[N_{ij}^{-1}\sum_{k=1}^{N_i}S_{ijk}(\bD^D_{ij}-\bmu^D_j)\{\epsilon_{ijk}+(\bD^D_{ij}-\bmu^D_j)\trans (
    \bh_j^D(\bX_{ik})-\bbeta^D)\}\right]\\
    &= \sum_{j=1}^J E[(\bD^D_{ij}-\bmu^D_j)(\bD^D_{ij}-\bmu^D_j)\trans]E[
    \bh_j^D(\bX_{ik})-\bbeta^D].
\end{align*}

For $E[\bh_j^D(\bX_{ik})]$, we observe that its non-zero entries satisfies
\begin{align*}
    E[h_j^D(\bX_{ik};d)] &= E[Y_{ijk}\mid Z_i=j-d+1] - E[Y_{ijk}\mid Z_i>j]  \\
    &= E[Y_{ijk}(j-d+1)] - E[Y_{ijk}(0)]\\
    &= \Delta(d),
\end{align*}
under the duration-specific treatment effect structure. Furthermore, $E[h_j^D(\bX_{ik};d)]$ does not vary by $j$. Therefore, it is easy to verify that $\bbeta^D = \bDelta^D$ is the solution to $E\bpsi_i^D(\bbeta^D,\bG_i) = 0$. It is also the unique solution because $E[(\bD^D_{ij}-\bmu^D_j)(\bD^D_{ij}-\bmu^D_j)\trans]$ is positive-definite for $j=J$ and positive semi-definite for $j< J$. (In the special case $P(Z=j)=0$ for some $j$, we can easily re-parameterize the functions to achieve uniqueness.)

For period-specific and saturated treatment effects, we can follow the same steps to prove that the estimating equation is unbiased.



\vspace{0.2in}
\noindent\textbf{Part II (Unbiased estimating equation given model misspecification)}

 For arbitrary limiting function $\underline\bG_i$, we have 
\[
E( \bpsi_i^*(\bDelta^*,\bG_i))-E( \bpsi_i^*(\bDelta^*,\underline\bG_i))=\bzero.
\]
This is because, under Assumption 2,
\begin{align*}
  E( \bpsi_i^*(\bDelta^*,\bG_i))&-E( \bpsi_i^*(\bDelta^*,\underline\bG_i))= E[(\bmD_i^*-\bmu_i^*)\trans\bmS_i\trans \bmS_i\{\underline\bG_i-\bG_{i}\}]\\
  &=E[E(\bmD_i^*-\bmu_i^*\mid N_i)\trans E\{\bmS_i\trans \bmS_i\{\underline\bG_i-\bG_{i}\}\mid N_i\}]=\bzero.
\end{align*}

\noindent\textbf{Part III (Consistency and asymptotic normality)}

 We first define
 \[
 \bmV^*=E\{(\bmD_i^*-\bmu_i^*)\trans\bmS_i\trans \bmS_i\{\bmD_i^*-\bmu_i^*\}\},
 \]
 and recall the definition
 \[
\widehat\bmV^*=\frac{1}{I}\sum_{i=1}^{I}(\bmS_i\bmD_i^*-\bmS_i\bmu_i^*)\trans(\bmS_i\bmD_i^*-\bmS_i\bmu_i^*).
\]
By law of large numbers, we obtain $\Vert\widehat\bmV^*-\bmV^*\Vert=o_p(1)$.

    Next, we show $E[\bpsi_i^*(\bDelta^*,\widehat{\bG}_i^{(m)})\mid (\bW_i)_{i\in B_m^C}]=0$.  We define the function
    \[
    h(r)=E[\bpsi_i^*(\bDelta^*,\underline\bG_i+r(\widehat{\bG}_i^{(m)}-\underline\bG_i))\mid (\bW_i)_{i\in B_m^C}],
    \]
    and clearly, $h(1)=E[\bpsi_i^*(\bDelta^*,\widehat{\bG}_i^{(m)})\mid (\bW_i)_{i\in B_m^C}]$ and $
    h(0)=E[\bpsi_i^*(\bDelta^*,\underline\bG_i)\mid (\bW_i)_{i\in B_m^C}]=E[\bpsi_i^*(\bDelta^*,\underline\bG_i)]=\bzero$
    because of the independence between different partitions of data. By Taylor expansion, for some value $\tilde{r}\in [0,1]$ 
    \begin{align*}
        &h(1)=h(0)+h'(0)+h(\tilde{r})''/2\\
        &=\bzero+\frac{\partial E[\bpsi\{\bDelta^*,\underline\bG_i+r(\widehat{\bG}_i^{(m)}-\underline\bG_i)\}\mid (\bW_i)_{i\in B_m^C}]}{\partial r}\bigg|_{r=0}+\frac{\partial^2 E[\bpsi\{\bDelta^*,\underline\bG_i+r(\widehat{\bG}_i^{(m)}-\underline\bG_i)\}\mid (\bW_i)_{i\in B_m^C}]}{2(\partial r)^2}\bigg|_{r=\tilde{r}}\\
        &=-E[(\bmD_i^*-\bmu_i^*)\trans\bmS_i\trans \bmS_i\{\widehat{\bG}_i^{(m\}}-\underline\bG_i)\mid (\bW_i)_{i\in B_m^C}]-\frac{\partial E[(\bmD_i^*-\bmu_i^*)\trans\bmS_i\trans \bmS_i\{\widehat{\bG}_i^{(m\}}-\underline\bG_i)\mid (\bW_i)_{i\in B_m^C}]}{2\partial r}\\
        &=\bzero,
    \end{align*}
    where the last equation is satisfied because $\bmD_i^*$ is independent of $\bX_{ik}$ given $N_i$, $\widehat{\bG}_i^{(m)}$ is non-stochastic given training data $(\bW_i)_{i\in B_m^C}$, and data in different partitions are independent. This derivation implies $E[\bpsi_i^*(\bDelta^*,\widehat{\bG}_i^{(m)})\mid (\bW_i)_{i\in B_m^C}]=0$ and hence $E[\bpsi_i^*(\bDelta^*,\widehat{\bG}_i^{(m)})]=0$.

    Using this result, we can obtain 
    \begin{align*}
        &E\left[\bigg\Vert\frac{1}{n}\sum_{i\in B_m}\bpsi_i^*(\bDelta^*,\widehat{\bG}_i^{(m)})-\frac{1}{n}\sum_{i\in B_m}\bpsi_i^*(\bDelta^*,\underline\bG_i)\bigg\Vert^2\right]\\
        &=E\left[E\left[\bigg\Vert\frac{1}{n}\sum_{i\in B_m}\bpsi_i^*(\bDelta^*,\widehat{\bG}_i^{(m)})-E[\bpsi_i^*(\bDelta^*,\widehat{\bG}_i^{(m)})]+E[\bpsi_i^*(\bDelta^*,\underline\bG_i)]-\frac{1}{n}\sum_{i\in B_m}\bpsi_i^*(\bDelta^*,\underline\bG_i)\bigg\Vert^2\biggm| (\bW_i)_{i\in B_m^C}\right]\right]\\
        &\leq n^{-1}E[E\{\Vert\bpsi_i^*(\bDelta^*,\widehat{\bG}_i^{(m)})-\bpsi_i^*(\bDelta^*,\underline\bG_i)\Vert^2\mid (\bW_i)_{i\in B_m^C}\}]
        \\
        &=n^{-1}E[E\{\Vert(\bmD_i^*-\bmu_i^*)\trans \bmS_i\trans \bmS_i\{\widehat{\bG}_i^{(m\}}-\underline\bG_i)\Vert^2\mid (\bW_i)_{i\in B_m^C}\}]\\
        &\leq n^{-1}CE[E\{\Vert\bmS_i\{\widehat{\bG}_i^{(m\}}-\underline\bG_i)\Vert^2\mid (\bW_i)_{i\in B_m^C}\}]\\
        &= n^{-1}C E[E\{\sum_{j=1}^J N_{ij}^{-1} \sum_{k=1}^{N_i} S_{ijk}(\widehat{g}_j^{(m)}(\bX_{ik})-\underline{g}_j(\bX_{ik}))^2\mid (\bW_i)_{i\in B_m^C}\}]\\
        &= n^{-1}\sum_{j=1}^J C E[E\{  \sum_{k=1}^{N_i} E[N_{ij}^{-1}S_{ijk}\mid N_i,(\bW_i)_{i\in B_m^C}\}]E[(\widehat{g}_j^{(m)}(\bX_{ik})-\underline{g}_j(\bX_{ik}))^2\mid N_i,(\bW_i)_{i\in B_m^C}\}]\mid (\bW_i)_{i\in B_m^C}\}]\\
        &= n^{-1}C \sum_{j=1}^J  E[(\widehat{g}_j^{(m)}(\bX_{ik})-\underline{g}_j(\bX_{ik}))^2]\\
        &=o(n^{-1}),\\
    \end{align*}
    where first inequality (third line) results from the independence of data across $i$, the fifth line uses the fact that $||(\bmD_i^*-\bmu_i^*)\trans \bmS_i\trans||^2 \le C$ for a constant $C$, the seventh line uses Assumption 3, the eighth line uses Assumption 1(2), and the last line results is implied by the convergence condition on nuisance function estimation. 
   By definition, $n=I/M$ and $M$ is a fixed integer, and thus we have $o(n)=o(I)$. Then, we can obtain that, by Chebyshev inequality,
    \[
    \frac{1}{M}\sum_{m=1}^{M}\frac{1}{n}\sum_{i\in B_m}\bpsi_i^*(\bDelta^*,\widehat{\bG}_i^{(m)})-\frac{1}{M}\sum_{m=1}^{M}\frac{1}{n}\sum_{i\in B_m}\bpsi_i^*(\bDelta^*,\underline\bG_i)=o_p(I^{-1/2}).
    \]
   By solving the estimating equation $\sum_{m=1}^{M}\sum_{i\in B_m}  \bpsi_i^*(\bbeta^*,\widehat\bG^{(m)}_i) = \bzero$, we have
\[
\widehat{\bbeta}^*=\left\{\sum_{m=1}^{M}\sum_{i\in B_m}(\bmS_i\bmD_i^*-\bmS_i\bmu_i^*)\trans(\bmS_i\bmD_i^*-\bmS_i\bmu_i^*)\right\}^{-1}\left\{\sum_{m=1}^{M}\sum_{i\in B_m}(\bmS_i\bmD_i^*-\bmS_i\bmu_i^*)\trans(\bmS_i\bY_i-\bmS_i\widehat\bG^{(m)}_i)\right\}.
\]
    Then, we have
    \begin{align*}
        &\widehat{\bbeta}^*-\bDelta^*=(\widehat\bmV^*)^{-1}\frac{1}{M}\sum_{m=1}^{M}\frac{1}{n}\sum_{i\in B_m}(\bmS_i\bmD_i^*-\bmS_i\bmu_i^*)\trans(\bmS_i\bY_i-\bmS_i\widehat{\bG}_i^{(m)})-\bDelta^*\\
        &=(\widehat\bmV^*)^{-1}\bigg(\frac{1}{M}\sum_{m=1}^{M}\frac{1}{n}\sum_{i\in B_m}(\bmS_i\bmD_i^*-\bmS_i\bmu_i^*)\trans(\bmS_i\bY_i-\bmS_i\widehat{\bG}_i^{(m)})-\frac{1}{I}\sum_{i=1}^{I}(\bmS_i\bmD_i^*-\bmS_i\bmu_i^*)\trans(\bmS_i\bmD_i^*-\bmS_i\bmu_i^*)\bDelta^*\bigg)\\
        &=(\widehat\bmV^*)^{-1}\bigg(\frac{1}{M}\sum_{m=1}^{M}\frac{1}{n}\sum_{i\in B_m}\bpsi_i^*(\bDelta^*,\widehat{\bG}_i^{(m)})\bigg)\\
        &=(\bmV^*+o_p(1))^{-1}\bigg(\frac{1}{M}\sum_{m=1}^{M}\frac{1}{n}\sum_{i\in B_m}\bpsi_i^*(\bDelta^*,\underline\bG_i)+o_p(I^{-1/2})\bigg)\\
        &=(\bmV^*+o_p(1))^{-1}\bigg(\frac{1}{I}\sum_{i=1}^{I}\bpsi_i^*(\bDelta^*,\underline\bG_i)+o_p(I^{-1/2})\bigg).
    \end{align*}
    By central limit theorem, we have 
    \[
    \frac{1}{I}\sum_{i=1}^{I}\bpsi_i^*(\bDelta^*,\underline\bG_i)-E\bpsi_i^*(\bDelta^*,\underline\bG_i)=O_p(I^{-1/2}).
    \]
    Since $\bmV^*$ is positive definite, we obtain
    \begin{align*}
       \widehat{\bbeta}^*-\bDelta^*&=(\bmV^*+o_p(1))^{-1}\bigg(\frac{1}{I}\sum_{i=1}^{I}\bpsi_i^*(\bDelta^*,\underline\bG_i)+o_p(I^{-1/2})\bigg) \\
       &=(\bmV^*+o_p(1))^{-1}(O_p(I^{-1/2})+o_p(I^{-1/2}))=O_p(I^{-1/2}).
    \end{align*}
    Furthermore, by central limit theorem, we know $I^{-1}\sum_{i=1}^{I}\bpsi_i^*(\bDelta^*,\underline\bG_i)$ convergence to normal distribution. Then, for
    \begin{align*}
        \sqrt{I}(\widehat{\bbeta}^*-\bDelta^*)=(\bmV^*)^{-1}\frac{1}{\sqrt I}\sum_{i=1}^{I}\bpsi_i^*(\bDelta^*,\underline\bG_i)+o_p(1),
    \end{align*}
    by Delta theorem, $\sqrt{I}(\widehat{\bbeta}^*-\bDelta^*)$ convergence to normal distribution $N(\bzero,\bSigma^*)$, where
    \[
    \bSigma^*=(\bmV^*)^{-1}E\{\bpsi_i^*(\bDelta^*,\underline\bG_i)\bpsi_i^*(\bDelta^*,\underline\bG_i)\trans\}(\bmV^*)^{-1}
    \]
\noindent\textbf{Part IV (Consistency of the variance estimator)}

    In this part, we show that $\widehat{\bSigma}^*$ from equation \eqref{var-est: 1} converges to $\bSigma^*$ in probability. Suppose $\widehat{\bpsi_i^*}=\bpsi_i^*(\widehat{\bbeta}^*,\widehat{\bG}_i^{(m)})$ and $\bpsi_i^*=\bpsi_i^*(\bDelta^*,\underline\bG_i)$, then we have 
    \begin{align*}
        \bigg\Vert\frac{1}{n}\sum_{i\in B_m}&\widehat{\bpsi_i^*}\widehat{\bpsi_i^*}\trans-\frac{1}{n}\sum_{i\in B_m}\bpsi_i^*{\bpsi_i^*}\trans\bigg\Vert\leq \frac{1}{n}\sum_{i\in B_m}\Vert\widehat{\bpsi_i^*}\widehat{\bpsi_i^*}\trans-\bpsi_i^*{\bpsi_i^*}\trans\Vert\\
        &\leq\frac{1}{n}\sum_{i\in B_m}\Vert(\widehat{\bpsi_i^*}-\bpsi_i^*)(\widehat{\bpsi_i^*}-\bpsi_i^*)\trans\Vert+\frac{2}{n}\sum_{i\in B_m}\Vert(\widehat{\bpsi_i^*}-\bpsi_i^*){\bpsi_i^*}\trans\Vert\\
        &\leq \frac{1}{n}\sum_{i\in B_m}\Vert(\widehat{\bpsi_i^*}-\bpsi_i^*)\Vert^2+2\bigg (\frac{1}{n}\sum_{i\in B_m}\Vert(\widehat{\bpsi_i^*}-\bpsi_i^*)\Vert^2 \bigg)^{1/2}\bigg (\frac{1}{n}\sum_{i\in B_m}\Vert\bpsi_i^*\Vert^2 \bigg)^{1/2}\\
        &\leq \bigg (\frac{1}{n}\sum_{i\in B_m}\Vert(\widehat{\bpsi_i^*}-\bpsi_i^*)\Vert^2 \bigg)^{1/2}\bigg\{\bigg (\frac{1}{n}\sum_{i\in B_m}\Vert(\widehat{\bpsi_i^*}-\bpsi_i^*)\Vert^2 \bigg)^{1/2}+2\bigg(\frac{1}{n}\sum_{i\in B_m}\Vert\bpsi_i^*\Vert^2 \bigg)^{1/2}\bigg\}.
    \end{align*}
    Because both $\widehat{\bpsi_i^*}$ and $\bpsi_i^*$ are vectors, the third inequality can be satisfied by applying Cauchy inequality. By using law of large number, we have that $n^{-1}\sum_{i\in B_m}\Vert\bpsi_i^*\Vert^2$ converges to $E\Vert\bpsi_i^*\Vert^2$ in probability. Moreover, we have
    \begin{align*}
        &\frac{1}{n}\sum_{i\in B_m}\Vert(\widehat{\bpsi_i^*}-\bpsi_i^*)\Vert^2=\frac{1}{n}\sum_{i\in B_m}\Vert(\bmD_i^*-\bmu_i^*)\trans\bmS_i\trans \bmS_i\{\bY_i-(\bmD_i^*-\bmu_i^*\}\widehat{\bbeta}^*-\widehat{\bG}_i^{(m)}-\bpsi_i^*\Vert^2\\
        &=\frac{1}{n}\sum_{i\in B_m}\Vert-(\bmD_i^*-\bmu_i^*)\trans\bmS_i\trans \bmS_i\{\bmD_i^*-\bmu_i^*\}(\widehat{\bbeta}^*-\bDelta^*)+(\bmD_i^*-\bmu_i^*)\trans\bmS_i\trans \bmS_i\{\bY_i-(\bmD_i^*-\bmu_i^*\}\bDelta^*-\widehat{\bG}_i^{(m)}-\bpsi_i^*\Vert^2
        \\
        &\leq \frac{1}{n}\sum_{i\in B_m}\Vert(\bmD_i^*-\bmu_i^*)\trans\bmS_i\trans \bmS_i\{\bmD_i^*-\bmu_i^*\}(\widehat{\bbeta}^*-\bDelta^*)\Vert^2+\frac{1}{n}\sum_{i\in B_m}\Vert\bpsi_i^*(\bDelta^*,\widehat{\bG}_i^{(m)})-\bpsi_i^*(\bDelta^*,\underline\bG_i)\Vert^2.
    \end{align*}
    The first term of the last line is bounded by 
    \[
    \bigg(\frac{1}{n}\sum_{i\in B_m}\Vert(\bmD_i^*-\bmu_i^*)\trans\bmS_i\trans \bmS_i\{\bmD_i^*-\bmu_i^*\}\Vert^2\bigg)\Vert(\widehat{\bbeta}^*-\bDelta^*)\Vert^2=O_p(1)O_p(I^{-1})=O_p(I^{-1}).
    \]
    The equation is satisfied because $\bmD_i^*$ is the matrix of indicator function and by law of large number, $n^{-1}\sum_{i\in B_m}\Vert(\bmD_i^*-\bmu_i^*)\trans\bmS_i\trans \bmS_i\{\bmD_i^*-\bmu_i^*\}\Vert^2$ converges to $E\Vert(\bmD_i^*-\bmu_i^*)\trans\bmS_i\trans \bmS_i\{\bmD_i^*-\bmu_i^*\}\Vert^2$. Therefore, we further have 
    \begin{align*}
        E[\Vert\bpsi_i^*(\bDelta^*,\widehat{\bG}_i^{(m)})-&\bpsi_i^*(\bDelta^*,\underline\bG_i)\Vert^2]\leq E[E\{\Vert\bpsi_i^*(\bDelta^*,\widehat{\bG}_i^{(m)})-\bpsi_i^*(\bDelta^*,\underline\bG_i)\Vert^2\mid (\bW_i)_{i\in B_m^C}\}]\\
        &=E[E\{\Vert(\bmD_i^*-\bmu_i^*)\trans \bmS_i\trans \bmS_i\{\widehat{\bG}_i^{(m\}}-\underline\bG_i)\Vert^2\mid (\bW_i)_{i\in B_m^C}\}]\\
         &\leq CE[E\{\Vert\widehat{\bG}_i^{(m)}-\underline\bG_i\Vert^2\mid (\bW_i)_{i\in B_m^C}\}]\\
        &=CE[\Vert\widehat{\bG}_i^{(m)}-\underline\bG_i\Vert^2]=o_p(1).
    \end{align*}
    Then by Markov inequality, we can have  
    \[
    \frac{1}{n}\sum_{i\in B_m}\Vert\bpsi_i^*(\bDelta^*,\widehat{\bG}_i^{(m)})-\bpsi_i^*(\bDelta^*,\underline\bG_i)\Vert^2=o_p(1).
    \]
    Therefore, we can obtain 
    \begin{align*}
         \bigg\Vert\frac{1}{n}\sum_{i\in B_m}&\widehat{\bpsi_i^*}\widehat{\bpsi_i^*}\trans-\frac{1}{n}\sum_{i\in B_m}\bpsi_i^*{\bpsi_i^*}\trans\bigg\Vert^2\\
         &\leq \bigg (\frac{1}{n}\sum_{i\in B_m}\Vert(\widehat{\bpsi_i^*}-\bpsi_i^*)\Vert^2 \bigg)\bigg\{\bigg (\frac{1}{n}\sum_{i\in B_m}\Vert(\widehat{\bpsi_i^*}-\bpsi_i^*)\Vert^2 \bigg)^{1/2}+2\bigg(\frac{1}{n}\sum_{i\in B_m}\Vert\bpsi_i^*\Vert^2 \bigg)^{1/2}\bigg\}^2\\
         &=o_p(1)(o_p(1)+O_p(1))^2=o_p(1).
    \end{align*}
    Furthermore, by law of large number, we have 
    \[
    \frac{1}{n}\sum_{i\in B_m}\bpsi_i^*{\bpsi_i^*}\trans-E\bpsi_i^*{\bpsi_i^*}\trans=O_p(I^{-1/2}).
    \]
    Therefore, we can obtain 
    \begin{align*}
        \bigg\Vert\frac{1}{n}\sum_{i\in B_m}\widehat{\bpsi_i^*}\widehat{\bpsi_i^*}\trans-E\bpsi_i^*{\bpsi_i^*}\trans\bigg\Vert&\leq \bigg\Vert\frac{1}{n}\sum_{i\in B_m}\widehat{\bpsi_i^*}\widehat{\bpsi_i^*}\trans-\frac{1}{n}\sum_{i\in B_m}\bpsi_i^*{\bpsi_i^*}\trans\bigg\Vert+ \bigg\Vert\frac{1}{n}\sum_{i\in B_m}\bpsi_i^*{\bpsi_i^*}\trans-E\bpsi_i^*{\bpsi_i^*}\trans\bigg\Vert\\
        &=o_p(1)+O_p(I^{-1/2})=o_p(1).
    \end{align*}
    Together with $\Vert\widehat\bmV-\bmV\Vert=o_p(1)$, and again, because $M$ is a fixed number, we have
    \begin{align*}
        \widehat{\bSigma}^*-\bSigma^*&=(\widehat\bmV^*)^{-1}\bigg\{\frac{1}{M}\sum_{m=1}^{M}\frac{1}{n}\sum_{i\in B_m}\bpsi_i^*(\widehat{\bbeta}^*,\widehat{\bG}_i^{(m)})\bpsi_i^*(\widehat{\bbeta}^*,\widehat{\bG}_i^{(m)})\trans\bigg\}(\widehat\bmV^*)^{-1}-\bSigma^*\\
        &=(\bmV^*+o_p(1))^{-1}(E\bpsi_i^*{\bpsi_i^*}\trans+o_p(1))(\bmV^*+o_p(1))^{-1}-\bSigma^*
        \\
        &=o_p(1).
    \end{align*}
Therefore, by Slutsky's Theorem, we obtain  $\sqrt{I}(\widehat{\bSigma}^*)^{-1}(\widehat{\bbeta}^*-\bDelta^*)\xrightarrow{d} N(\bzero,\bmI)$. 

\vspace{0.2in}
\noindent\textbf{Part V (Minimal asymptotic variance under certain conditions)}

In this part, we prove that, under the correct model specification, if $\underline{g}_j(\bX_{ik}) = E[Y_{ijk}|\bX_{ik}]$ and $Y_{ijk}$ is independent of $\bX_{ik'}$ for $k\ne k'$ given $\bX_{ik}$, $\widehat{\bbeta}^* $ achieves the smallest asymptotic variance. 
Without loss of generality, we assume $\bX_{ik}$ includes $N_i$ as a covariate.
Specifically, for every limit function $\underline\bG_i$, we need to prove 
\[
(\bmV^*)^{-1}E\{\bpsi_i^*(\bDelta^*,\underline\bG_i)\bpsi_i^*(\bDelta^*,\underline\bG_i)\trans\}(\bmV^*)^{-1}-(\bmV^*)^{-1}E\{\bpsi_i^*(\bDelta^*,\bG_i)\bpsi_i^*(\bDelta^*,\bG_i)\trans\}(\bmV^*)^{-1}\succeq \bzero,
\]
where $\succeq$ represents the matrix is semi-positive definite. It is equivalent to proving 
\[
\Var\{\bpsi_i^*(\bDelta^*,\underline\bG_i)\}-\Var\{\bpsi_i^*(\bDelta^*,\bG_i)\}\succeq \bzero,
\]
as $E\bpsi_i^*(\bDelta^*,\underline\bG_i)=E\bpsi_i^*(\bDelta^*,\bG_i)=\bzero$. We can derive 
\begin{align*}
   &\Var\{\bpsi_i^*(\bDelta^*,\underline\bG_i)\}-\Var\{\bpsi_i^*(\bDelta^*,\bG_i)\}=\Var\{\bpsi_i^*(\bDelta^*,\underline\bG_i)-\bpsi_i^*(\bDelta^*,\bG_i)\}+\\
   &\Cov\{\bpsi_i^*(\bDelta^*,\bG_i),\bpsi_i^*(\bDelta^*,\underline\bG_i)-\bpsi_i^*(\bDelta^*,\bG_i)\}+\Cov\{\bpsi_i^*(\bDelta^*,\underline\bG_i)-\bpsi_i^*(\bDelta^*,\bG_i),\bpsi_i^*(\bDelta^*,\bG_i)\}.
\end{align*}
Therefore, we only need to prove
\[
\Cov\{\bpsi_i^*(\bDelta^*,\bG_i),\bpsi_i^*(\bDelta^*,\underline\bG_i)-\bpsi_i^*(\bDelta^*,\bG_i)\}=E[\bpsi_i^*(\bDelta^*,\bG_i)\{\bpsi_i^*(\bDelta^*,\underline\bG_i)-\bpsi_i^*(\bDelta^*,\bG_i)\}\trans]=\bzero.
\]
We have 
\begin{equation}
\label{equ::minimal}
    \begin{aligned}
    &E[\bpsi_i^*(\bDelta^*,\bG_i)\{\bpsi_i^*(\bDelta^*,\underline\bG_i)-\bpsi_i^*(\bDelta^*,\bG_i)\}\trans]
    \\&=E[(\bmD_i^*-\bmu_i^*)\trans\bmS_i\trans \bmS_i\{(\bY_i-(\bmD_i^*-\bmu_i^*)\bDelta^*-\bG_i)\}(\bG_i-\underline\bG_i)\trans\bmS_i\trans \bmS_i\{\bmD_i^*-\bmu_i^*\}]
    \\&=E\left[\sum_{j=1}^J(\bD^*_{ij} - E[\bD^*_{ij}]) \frac{1}{N_{ij}}\sum_{k=1}^{N_i} S_{ijk}\epsilon_{ijk}'\sum_{j'=1}^J(\bD^*_{ij'} - E[\bD^*_{ij'}])\trans \frac{1}{N_{ij'}}\sum_{k'=1}^{N_i} S_{ij'k'} \left\{\underline g_j(\bX_{ik'})-g_j(\bX_{ik'})\right\}\right],
\end{aligned}
\end{equation}
where 
\[
\epsilon_{ijk}'=Y_{ijk}-(\bD^*_{ij} - E[\bD^*_{ij}])\trans\bDelta^* -g_j(\bX_{ik}).
\]
If the partial linear model in the main paper is correctly specified, we have $E[\epsilon_{ijk}'\mid Z_i,\bX_{ik}]=0$. Furthermore, $\bG_i$ and $\underline\bG_i$ are the function of $\bX_{ik'}$ for $k'\in {1,...,N_i}$. If $k=k'$, then we have 
\[
E\{\epsilon_{ijk}'\mid Z_i,\bX_{ik}\}(g_j(\bX_{ik})-\underline g_j(\bX_{ik}))=0.
\]
If $k\neq k'$, then under the condition that $Y_{ijk}$ is independent of $\bX_{ik'}$ for $k\ne k'$ given $\bX_{ik}$, we obtain
\[
E\{\epsilon_{ijk}'\mid Z_i,\bX_{ik}\}E\{(g_j(\bX_{ik'})-\underline g_j(\bX_{ik'})\mid \bX_{ik}\}=0.
\]
Denote $\bmA^*_{jj'}=(\bD^*_{ij} - E[\bD^*_{ij}])(\bD^*_{ij'} - E[\bD^*_{ij'}])\trans$, then, equation \ref{equ::minimal} is
\begin{align*}
    &E\left[\sum_{j=1}^J(\bD^*_{ij} - E[\bD^*_{ij}]) \frac{1}{N_{ij}}\sum_{k=1}^{N_i} S_{ijk}\epsilon_{ijk}'\sum_{j'=1}^J(\bD^*_{ij'} - E[\bD^*_{ij'}])\trans \frac{1}{N_{ij'}}\sum_{k'=1}^{N_i} S_{ij'k'} \left\{\underline g_j(\bX_{ik'})-g_j(\bX_{ik'})\right\}\right]\\
    &=E\left[\sum_{j=1}^J\sum_{j'=1}^J\bmA^*_{jj'} \frac{1}{N_{ij}N_{ij'}}\sum_{k=1}^{N_i} \sum_{k'=1}^{N_i}S_{ijk}S_{ij'k'}\epsilon_{ijk}'   \left\{\underline g_j(\bX_{ik'})-g_j(\bX_{ik'})\right\}\right]\\
    &=E\left[\sum_{j=1}^J\sum_{j'=1}^J\bmA^*_{jj'} \frac{1}{N_{ij}N_{ij'}}\sum_{k=1}^{N_i} S_{ijk}S_{ij'k}\epsilon_{ijk}'   \left\{\underline g_j(\bX_{ik})-g_j(\bX_{ik})\right\}\right]
    \\
    &+E\left[\sum_{j=1}^J\sum_{j'=1}^J\bmA^*_{jj'} \frac{1}{N_{ij}N_{ij'}}\sum_{k\neq k'} S_{ijk}S_{ij'k'}\epsilon_{ijk}'   \left\{\underline g_j(\bX_{ik'})-g_j(\bX_{ik'})\right\}\right]\\
    &=E\left[\sum_{j=1}^J\sum_{j'=1}^J\bmA^*_{jj'} E\left\{\frac{1}{N_{ij}N_{ij'}}\sum_{k=1}^{N_i} S_{ijk}S_{ij'k}\mid N_i\right\}E\{\epsilon_{ijk}'\mid Z_i,\bX_{ik}\}   \left\{\underline g_j(\bX_{ik})-g_j(\bX_{ik})\right\}\right]
    \\
    &+E\left[\sum_{j=1}^J\sum_{j'=1}^J\bmA^*_{jj'} E\left\{\frac{1}{N_{ij}N_{ij'}}\sum_{k\neq k'} S_{ijk}S_{ij’k'}\mid N_i\right\}E\{\epsilon_{ijk}'\mid Z_i,\bX_{ik}\}E\{(g_j(\bX_{ik'})-\underline g_j(\bX_{ik'})\mid \bX_{ik}\}\right]
    \\
    &=\bzero.
\end{align*}
For individual-randomized trial when $N_i=1$, equation \ref{equ::minimal} can be simplified as
\begin{align*}
    E\Bigg[\sum_{j=1}^J\sum_{j'=1}^J\bmA^*_{jj'} S_{ijk}&S_{ij'k}\epsilon_{ijk}' \left\{\underline g_j(\bX_{ik})-g_j(\bX_{ik})\right\}\Bigg]
    \\&=E\left[\sum_{j=1}^J\sum_{j'=1}^J\bmA^*_{jj'}   E[S_{ijk} S_{ij'k}]E[\epsilon_{ijk}'\mid Z_i,\bX_{ik}] \left\{\underline g_j(\bX_{ik})-g_j(\bX_{ik})\right\}\right]
    =0.
\end{align*}
Along with Assumptions 2 and 3 in the main paper, we have
\[
E[\bpsi_i^*(\bDelta^*,\bG_i)\{\bpsi_i^*(\bDelta^*,\underline\bG_i)-\bpsi_i^*(\bDelta^*,\bG_i)\}\trans]=\bzero.
\]
Hence, we have
\[
\Var\{\bpsi_i^*(\bDelta^*,\underline\bG_i)\}-\Var\{\bpsi_i^*(\bDelta^*,\bG_i)\}=\Var\{\bpsi_i^*(\bDelta^*,\underline\bG_i)-\bpsi_i^*(\bDelta^*,\bG_i)\}\succeq \bzero,
\]
which means when the nuisance is correctly specified, the estimator achieves the smallest asymptotic variance. 

\subsection{Proof for Theorem 2}
In this proof, we first establish pointwise and uniform convergence of the estimating function. Then, we establish the consistency and asymptotic normality of our estimator. Finally, we prove that the QIF estimator never reduces efficiency. 

Recall the definition
\begin{align*}
\bpsi_I^*(\bbeta^*,\widehat{\bG}) = \frac{1}{I}\sum_{i=1}^I\widetilde{\bpsi}^*_I\{\bbeta^*,\widehat{\bG}_i^{(m\}}) = \frac{1}{I}\sum_{i=1}^I\left(\begin{array}{c}
  (\bmD_i^* -\bmu^* )\trans \bmS_i\trans \bmQ_{i1} \bmS_i 
    \left\{\bY_i - (\bmD_i^* - \bmu^*) \bbeta^* - \widehat{\bG}_i^{(m)} \right\}   \\
    \vdots \\
  (\bmD_i^* - \bmu^*)\trans \bmS_i\trans \bmQ_{iL} \bmS_i 
    \left\{\bY_i - (\bmD_i^* - \bmu^*) \bbeta^* - \widehat{\bG}_i^{(m)} \right\}  
\end{array}\right).
\end{align*}
To simplify notation, we further denote
\[
\bmH_i^*=
\begin{pmatrix}
 (\bmD_i^*-\bmu^*)\trans \bmS_i\trans\bmQ_{i1}\bmS_i\\
\vdots \\
 (\bmD_i^*-\bmu^*)\trans \bmS_i\trans\bmQ_{iL}\bmS_i
\end{pmatrix}.
\]

According to \cite{qu2000improving} and \cite{hansen1982large}, the estimation is equivalent to solving the estimating equation
\[
 \bPsi_I^*(\bbeta^*,\widehat\bG)=\frac{\partial \bpsi^*_I\{\bbeta^*,\widehat\bG\}}{\partial (\bbeta^*)\trans}\{\bmC^*_I\{\bbeta^*,\widehat\bG\}\}^{-1}\bpsi^*_I\{\bbeta^*,\widehat\bG\}=\bzero.
\]

\noindent\textbf{Part I (Point convergence of estimating function)}

We first notice
\[
\frac{\partial \bpsi_I^*}{\partial \bbeta^*}=-\frac{1}{I} \sum_{i=1}^I
\begin{pmatrix}
 (\bmD_i^*-\bmu^*)\trans \bmS_i\trans\bmQ_{i1}\bmS_i\{\bmD_i^*-\bmu^*\}\\
\vdots \\
 (\bmD_i^*-\bmu^*)\trans \bmS_i\trans\bmQ_{iL}\bmS_i\{\bmD_i^*-\bmu^*\}
\end{pmatrix}
\]
is invariant to $\bbeta^*$ and $\bG$, and thus the  law of large number implies $\partial \bpsi_I^*/\partial \bbeta^*$ converges in probability to $E[\partial \bpsi_I^*/\partial \bbeta^*]$.


Then, for any $\bbeta^*$ in a compact set, we have 
\[E[\widetilde{\bpsi}^*_I\{\bbeta^*,\widehat{\bG}_i^{(m\}})\mid (\bW_i)_{i\in B_m^C}]-E[\widetilde{\bpsi}^*_I\{\bbeta^*,\underline\bG_i\} ]=E[E\{\bmH_i^*(\widehat{\bG}_i^{(m)}-\underline\bG_i))\}\mid (\bW_i)_{i\in B_m^C}]=\bzero.
 \]
Following the same proof in Parts I, II, and III of Section B.1, we have $E[\bpsi_I^*(\bDelta^*,\underline\bG_i)] = 0$ and, 
\begin{align*}
& \bpsi_I^*(\bbeta^*,\widehat {\bG})-\bpsi_I^*(\bbeta^*,\underline\bG)=o_p(I^{-1/2}).\\
&\bmC_I^*(\bbeta^*,\widehat\bG)-\bmC_I^*(\bbeta^*,\underline\bG)=o_p(1).
\end{align*}
Defining
\[
\bPhi^*(\bbeta^*,\underline\bG)=E\left[\frac{\partial \bpsi_I^*}{\partial (\bbeta^*)\trans}\right][E\{\bmC_I^*(\bbeta^*,\underline\bG)\}]^{-1}E\bpsi_{I}^*(\bbeta^*,\underline\bG),
\]
the above facts imply 
$\bPhi^*(\bDelta^*,\underline\bG)=\bzero$, i.e., unbiased estimating function, and
\begin{align*}
    &\bPsi_I^*(\bbeta^*,\widehat\bG)-\bPhi^*(\bbeta^*,\underline\bG)
    \\&=\frac{\partial \bpsi_I^*}{\partial (\bbeta^*)\trans}\{\bmC_I^*(\bbeta^*,\widehat {\bG})\}^{-1}\bpsi_I^*(\bbeta^*,\widehat {\bG})-E\left[\frac{\partial \bpsi_I^*}{\partial (\bbeta^*)\trans}\right][E\{\bmC_I^*(\bbeta^*,\underline\bG)\}]^{-1}E\bpsi_{I}^*(\bbeta^*,\underline\bG)\\
    &=\left(E\left[\frac{\partial \bpsi_I^*}{\partial (\bbeta^*)\trans}\right]+o_p(1)\right)[E\bmC_I^*(\bbeta^*,\underline\bG)+o_p(1)]^{-1}(E[\bpsi_{I}^*(\bbeta^*,\underline\bG)]+o_p(1))-\bPhi^*\{\bbeta^*,\underline\bG\}
    \\
    &=o_p(1).
\end{align*}
Therefore, we proved the point consistency of $\bPsi_I^*(\bbeta^*,\widehat {\bG})$.

\vspace{0.2in}
\noindent\textbf{Part II (Uniform convergence of estimating function)}

To obtain the uniform convergence, we show the stochastic equicontinuity of $\bPsi_I^*(\bbeta^*,\widehat {\bG})$. The simplest way is to show the Lipschitz continuity.  To simplify the illustration of this part, because $o(n)=o(I)$ and $M$ is a fixed number, we assume $\widehat\bG$ is estimated from training data $(\bW_i)_{\text{train}}$. Then, conditional on $(\bW_i)_{\text{train}}$, for $\delta>0$ and $(\bbeta^*,\widetilde{\bbeta}^*)$ from compact set $\mathscr{B}$ with $\Vert\widetilde{\bbeta}^*-\bbeta^*\Vert<\delta$, we denote
\begin{align*}
   \bM_1=\frac{1}{I}\sum_{i=1}^I\bmH_i^*(\bmD_i^*-\bmu^*)(\widetilde{\bbeta}^*-\bbeta^*),
\end{align*}
and
\begin{align*}
   \bmM_2=&\frac{1}{I}\sum_{i=1}^I\bmH_i^*(\bmD_i^*-\bmu^*)(\widetilde{\bbeta}^*-\bbeta^*)\bpsi_I^*(\bbeta^*,\widehat {\bG})\trans+ \frac{1}{I}\sum_{i=1}^I\bpsi_I^*(\bbeta^*,\widehat {\bG})\{\bmH_i^*(\bmD_i^*-\bmu^*)(\widetilde{\bbeta}^*-\bbeta^*)\}\trans
  \\
  &+\frac{1}{I}\sum_{i=1}^I\bmH_i^*(\bmD_i^*-\bmu^*)(\widetilde{\bbeta}^*-\bbeta^*)\{\bmH_i^*(\bmD_i^*-\bmu^*)(\widetilde{\bbeta}^*-\bbeta^*)\}\trans
\end{align*}
Therefore, we have that, 
\begin{align*}
    &\Vert\{\bmC_I^*(\widetilde{\bbeta}^*,\widehat {\bG})\}^{-1}-\{\bmC_I^*(\bbeta^*,\widehat {\bG})\}^{-1}\Vert=\Vert\{\bmC_I^*(\bbeta^*,\widehat {\bG})-\bmM_2\}^{-1}-\{\bmC_I^*(\bbeta^*,\widehat {\bG})\}^{-1}\Vert\\
    &=\Vert\{\bmC_I^*(\bbeta^*,\widehat {\bG})\}^{-1}+\{\bmC_I^*(\bbeta^*,\widehat {\bG})\}^{-1}\bmM_2\{\bmC_I^*(\bbeta^*,\widehat {\bG})-\bmM_2\}^{-1}-\{\bmC_I^*(\bbeta^*,\widehat {\bG})\}^{-1}\Vert
    \\
    &=\Vert\{\bmC_I^*(\bbeta^*,\widehat {\bG})\}^{-1}\bmM_2\{\bmC_I^*(\bbeta^*,\widehat {\bG})-\bmM_2\}^{-1}\Vert
    \\
    &\leq\Vert\{\bmC_I^*(\bbeta^*,\widehat {\bG})\}^{-1}\Vert\Vert\bmM_2\Vert\Vert\{\bmC_I^*(\widetilde{\bbeta}^*,\widehat {\bG})\}^{-1}\Vert
    \\
    &\leq \bL_1\delta,
\end{align*}
for
\begin{align*}
    \bL_1&=\Vert\{\bmC_I^*(\bbeta^*,\widehat {\bG})\}^{-1}\Vert\Vert\{\bmC_I^*(\widetilde{\bbeta}^*,\widehat {\bG})\}^{-1}\Vert \\
    &\times\left[2\left\Vert\frac{1}{I}\sum_{i=1}^I\bmH_i^*(\bmD_i^*-\bmu^*)\bpsi_I^*(\bbeta^*,\widehat {\bG})\trans\right\Vert+\left\Vert\frac{1}{I}\sum_{i=1}^I\bmH_i^*(\bmD_i^*-\bmu^*)\{\bmH_i^*(\bmD_i^*-\bmu^*)\}\trans \right\Vert\delta\right],
\end{align*}
following the previous results, $\bL_1$ converges to
\begin{align*}
    \Vert\{E\bmC_I^*(\bbeta^*,\widehat\bG)\}^{-1}&\Vert\Vert\{E\bmC_I^*(\widetilde{\bbeta}^*,\widehat\bG)\}^{-1}\Vert \\
    &\times
    (2\Vert E[\bmH_i^*(\bmD_i^*-\bmu^*)\bpsi_I^*(\bbeta^*,\widehat\bG)\trans]\Vert+\Vert E[\bmH_i^*(\bmD_i^*-\bmu^*)\{\bmH_i^*(\bmD_i^*-\bmu^*)\}\trans]\Vert \delta),
\end{align*}
 By definition, $\bmC_I^*(\bbeta^*,\widehat\bG)$ and $\bpsi_I^*(\bbeta^*,\widehat\bG)$ are continuous function of $\bbeta^*$, and $\bbeta^*,\widetilde{\bbeta}^*$ are from compact set. Therefore, conditional on $(\bW_i)_{\text{train}}$, the formula above is bounded for $\delta<1$. Then we have,  
\begin{align*}
    &\Vert\bPsi_I^*(\widetilde{\bbeta}^*,\widehat {\bG})-\bPsi_I^*(\bbeta^*,\widehat {\bG})\Vert \le\left\Vert\frac{\partial \bpsi_I^*}{\partial (\bbeta^*)\trans}\right\Vert \Vert \{\bmC_I^*(\widetilde{\bbeta}^*,\widehat {\bG})\}^{-1}\bpsi_I^*(\widetilde{\bbeta}^*,\widehat {\bG})-\{\bmC_I^*(\bbeta^*,\widehat {\bG})\}^{-1}\bpsi_I^*(\bbeta^*,\widehat {\bG})\Vert\\
    &=\left\Vert\frac{\partial \bpsi_I^*}{\partial (\bbeta^*)\trans}\right\Vert\Vert \{\bmC_I^*(\widetilde{\bbeta}^*,\widehat {\bG})\}^{-1}\{\bpsi_I^*(\widetilde{\bbeta}^*,\widehat {\bG})-\bpsi_I^*(\bbeta^*,\widehat {\bG})\}+[\{\bmC_I^*(\widetilde{\bbeta}^*,\widehat {\bG})\}^{-1}-\{\bmC_I^*(\bbeta^*,\widehat {\bG})\}^{-1}]\bpsi_I^*(\bbeta^*,\widehat {\bG})\Vert
    \\
    &\leq\left\Vert\frac{\partial \bpsi_I^*}{\partial (\bbeta^*)\trans}\right\Vert[\Vert \{\bmC_I^*(\widetilde{\bbeta}^*,\widehat {\bG})\}^{-1}\Vert\Vert\bM_1\Vert+\Vert\{\bmC_I^*(\widetilde{\bbeta}^*,\widehat {\bG})\}^{-1}-\{\bmC_I^*(\bbeta^*,\widehat {\bG})\}^{-1}\Vert\Vert\bpsi_I^*(\bbeta^*,\widehat {\bG})\Vert]
    \\
    &\leq\left\Vert\frac{\partial \bpsi_I^*}{\partial (\bbeta^*)\trans}\right\Vert\left[\Vert \{\bmC_I^*(\widetilde{\bbeta}^*,\widehat {\bG})\}^{-1}\Vert\left\Vert\frac{1}{I}\sum_{i=1}^I\bmH_i^*(\bmD_i^*-\bmu^*)\right\Vert+\bL_1\Vert\bpsi_I^*(\bbeta^*,\widehat {\bG})\Vert\right]\delta\\
    &\triangleq \bL_2\delta,
\end{align*}
for $I\to\infty$, with high probability, $\bL_2$ is bounded by some constant $c_1$ for $\delta<1$. Therefore, conditional on $(\bW_i)_{\text{train}}$, for $I\to\infty$, for every $\epsilon>0$, there exists $\delta<\min (\epsilon/c_1,1)$, we have 
\[
 P(  \sup_{\bbeta^*,\widetilde{\bbeta}^*,\Vert\widetilde{\bbeta}^*-\bbeta^*\Vert<\delta}\Vert\bPsi_I^*(\widetilde{\bbeta}^*,\widehat {\bG}) - \bPsi_I^*(\bbeta^*,\widehat {\bG})\Vert > \epsilon \mid (\bW_i)_{\text{train}}) \to 0.
\]
Then we have 
\begin{align*}
   P(\sup_{\bbeta^*,\widetilde{\bbeta}^*,\Vert\widetilde{\bbeta}^*-\bbeta^*\Vert<\delta}  \Vert\bPsi_I\{\widetilde{\bbeta}^*&,\widehat {\bG}\} - \bPsi_I^*(\bbeta^*,\widehat {\bG})\Vert > \epsilon )
   \\
   &\leq E[P( \sup_{\bbeta^*,\widetilde{\bbeta}^*,\Vert\widetilde{\bbeta}^*-\bbeta^*\Vert<\delta} \Vert\bPsi_I^*(\widetilde{\bbeta}^*,\widehat {\bG}) - \bPsi_I^*(\bbeta^*,\widehat {\bG})\Vert > \epsilon \mid (\bW_i)_{\text{train}})] \to 0, 
\end{align*}
since $P(  \sup_{\bbeta^*,\widetilde{\bbeta}^*,\Vert\widetilde{\bbeta}^*-\bbeta^*\Vert<\delta}\Vert\bPsi_I^*(\widetilde{\bbeta}^*,\widehat {\bG}) - \bPsi_I^*(\bbeta^*,\widehat {\bG})\Vert > \epsilon \mid (\bW_i)_{\text{train}})$ is uniformly integrable. Therefore, we proved the stochastic equicontinuity of $\bPsi_I^*(\bbeta^*,\widehat {\bG})$. Following Arzelà–Ascoli theorem, we have that, for
 $I\to\infty$,
\[
\sup_{\bbeta^*} \Vert\bPsi_I^*(\bbeta^*,\widehat\bG)-\bPhi^*(\bbeta^*,\underline\bG)\Vert\stackrel{p}{\rightarrow} 0.
\]
Then 5.9 theorem in \cite{van2000asymptotic} and \cite{hansen1982large} imply $\widehat{\bbeta}^*_{\textup{QIF}}$ converges to $\bDelta^*$ in probability.

\vspace{0.2in}
\noindent\textbf{Part III (Asymptotic normality)}

Following the proof in Part III of Section B.1, along with the consistency of $\widehat{\bbeta}^*_{\textup{QIF}}$, we have 
\[
 \bmC_I^*(\widehat{\bbeta}^*_{\textup{QIF}},\widehat\bG)-\bmC_I^*(\bDelta^*,\underline\bG)=o_p(1).
\]
By law of large number, $\bmC_I^*(\bDelta^*,\underline\bG)$ converges to $E\bmC_I^*(\bDelta^*,\underline\bG)$ in probability, and thus we have $\bmC_I^*(\widehat{\bbeta}^*_{\textup{QIF}},\widehat\bG)-E\bmC_I^*(\bDelta^*,\underline\bG)=o_p(1)$. 

 Then, we analyze the convergence rate of $\widehat{\bbeta}^*_{\textup{QIF}}$. We already have that
    \[
    \bpsi_I^*(\bDelta^*,\widehat{\bG})-\bpsi_I^*(\bDelta^*,\underline\bG)=\frac{1}{I}\sum_{i=1}^{I}\widetilde{\bpsi}^*_I\{\bDelta^*,\widehat{\bG}_i^{(m\}})-\frac{1}{I}\sum_{i=1}^{I}\widetilde{\bpsi}^*_I\{\bDelta^*,\underline\bG_i\}=o_p(I^{-1/2}).
    \]
Then we have
\begin{align*}
    &\sqrt{I}(\widehat{\bbeta}^*_{\textup{QIF}}-\bDelta^*)
    \\&=\left(-\frac{\partial \bpsi_I^*}{\partial (\bbeta^*)\trans}\{\bmC_I^*(\widehat{\bbeta}^*_{\textup{QIF}},\widehat\bG)\}^{-1}\frac{\partial \bpsi_I^*}{\partial \bbeta^*}\right)^{-1}\frac{\partial \bpsi_I^*}{\partial (\bbeta^*)\trans}\{\bmC_I^*(\widehat{\bbeta}^*_{\textup{QIF}},\widehat\bG)\}^{-1}\frac{1}{\sqrt{I}}\sum_{i=1}^I\bmH_i^*(\bY_i-\widehat{\bG}_i^{(m)})-\sqrt{I}\bDelta^*\\
   &= \left(-\frac{\partial \bpsi_I^*}{\partial (\bbeta^*)\trans}\{\bmC_I^*(\widehat{\bbeta}^*_{\textup{QIF}},\widehat\bG)\}^{-1}\frac{\partial \bpsi_I^*}{\partial \bbeta^*}\right)^{-1}\frac{\partial \bpsi_I^*}{\partial (\bbeta^*)\trans}\{\bmC_I^*(\widehat{\bbeta}^*_{\textup{QIF}},\widehat\bG)\}^{-1}\sqrt{I}\bpsi^*_I\{\bDelta^*,\widehat{\bG}\}
    \\
    &=\left(-E\left[\frac{\partial \bpsi_I^*}{\partial (\bbeta^*)\trans}\right][E\{\bmC_I^*(\bDelta^*,\underline\bG)\}]^{-1}E\left[\frac{\partial \bpsi_I^*}{\partial \bbeta^*}\right]\right)^{-1}E\left[\frac{\partial \bpsi_I^*}{\partial (\bbeta^*)\trans}\right][E\{\bmC_I^*(\bDelta^*,\underline\bG)\}]^{-1}\sqrt{I}\bpsi_I^*(\bDelta^*,\underline\bG)+o_p(1)\\
    &=\left(-E\left[\frac{\partial \bpsi_I^*}{\partial (\bbeta^*)\trans}\right][E\{\bmC_I^*(\bDelta^*,\underline\bG)\}]^{-1}E\left[\frac{\partial \bpsi_I^*}{\partial \bbeta^*}\right]\right)^{-1}\bpsi_E(\bDelta^*,\underline\bG)+o_p(1)
\end{align*}
where
\[
\bpsi_E(\bDelta^*,\underline\bG)=\sqrt{I}E\left[\frac{\partial \bpsi_I^*}{\partial (\bbeta^*)\trans}\right][E\{\bmC_I^*(\bDelta^*,\underline\bG)\}]^{-1}\bpsi_I^*(\bDelta^*,\underline\bG),
\]
and we have $E[\bpsi_E(\bDelta^*,\underline\bG)]=\bzero$. Therefore, we obtain that $\widehat{\bbeta}^*_{\textup{QIF}}-\bDelta^*=O_p(I^{-1/2})$ and $\sqrt{I}(\widehat{\bbeta}^*_{\textup{QIF}}-\bDelta^*)$ converges to normal distribution with mean $\bzero$ and variance 
\[
\bSigma^*_{\textup{QIF}}=\left(E\left[\frac{\partial \bpsi_I^*}{\partial (\bbeta^*)\trans}\right][E\{\bmC_I^*(\bDelta^*,\underline\bG)\}]^{-1}E\left[\frac{\partial \bpsi_I^*}{\partial \bbeta^*}\right]\right)^{-1}.
\]
By $\bmC_I^*(\widehat{\bbeta}^*_{\textup{QIF}},\widehat\bG)-E\bmC_I^*(\bDelta^*,\underline\bG)=o_p(1)$, it is straightfoward that the variance estimator $\widehat\bSigma^*_{\textup{QIF}}$ in Equation~\eqref{var-est: QIF} converges to $\bSigma^*_{\textup{QIF}}$ in probability. Then, by Slutsky's Theorem, we can have that $\sqrt{I}\widehat{\bSigma}^{-1}_{\textup{QIF}}(\widehat{\bbeta}^*_{\textup{QIF}}-\bDelta^*)\xrightarrow{d} N(\bzero,\bmI)$.

\vspace{0.2in}
\noindent\textbf{Part IV (No variance increase by QIF)}

Finally, we prove that using QIF methods leads to a smaller variance compared to basic estimators. Without loss of generality, We consdier use two correlation matrices for QIF. These two matrices construct two estimating equations for the extended score function $(\bpsi_1\trans,\bpsi_2\trans)\trans$. Then, the asymptotic variance of the basic method is
\[
\bSigma_1=\left(E(\dot{\bpsi_1}\trans)[E\{\bpsi_1\bpsi_1\trans\}]^{-1}E\dot\bpsi_1\right)^{-1},
\]
where $\dot{\bpsi_1}=E[\partial \bpsi_1/\partial \bbeta^*]$. The asymptotic variance of the QIF is
\[
\bSigma_2=\left(E\begin{pmatrix}
 \dot{\bpsi_1}\\
 \dot{\bpsi_2}
\end{pmatrix}\trans E\begin{pmatrix}
 \bpsi_1\bpsi_1\trans&\bpsi_1\bpsi_2\trans\\
 \bpsi_2\bpsi_1\trans&\bpsi_2\bpsi_2\trans
\end{pmatrix}^{-1}E\begin{pmatrix}
 \dot{\bpsi_1}\\
 \dot{\bpsi_2}
\end{pmatrix}\right)^{-1}.
\]
To prove $\bSigma_1-\bSigma_2\succeq 0$, we only need to prove $\bSigma_2^{-1}-\bSigma_1^{-1}\succeq 0$. By using Schur Complement for
\[
E\begin{pmatrix}
 \bpsi_1\bpsi_1\trans&\bpsi_1\bpsi_2\trans\\
 \bpsi_2\bpsi_1\trans&\bpsi_2\bpsi_2\trans
\end{pmatrix}\triangleq E\begin{pmatrix}
 A&B\\
 C&D
\end{pmatrix},
\]
we have
\[
E\begin{pmatrix}
 \bpsi_1\bpsi_1\trans&\bpsi_1\bpsi_2\trans\\
 \bpsi_2\bpsi_1\trans&\bpsi_2\bpsi_2\trans
\end{pmatrix}^{-1}=
E\begin{pmatrix}
 A^{-1}+A^{-1}BS^{-1}CA^{-1}&-A^{-1}BS^{-1}\\
 -S^{-1}CA^{-1}&S^{-1}
\end{pmatrix},
\]
where $S=D-CA^{-1}B$. Then, notice $B=C\trans$,
\begin{align*}
    &\bSigma_2^{-1}-\bSigma_1^{-1}=\\
    &\dot{\bpsi_1}\trans (A^{-1}+A^{-1}BS^{-1}CA^{-1})\dot{\bpsi_1}-\dot{\bpsi_1}\trans A^{-1}BS`^{-1}\dot{\bpsi_2}-\dot{\bpsi_2}\trans S^{-1}CA^{-1}\dot{\bpsi_1}+\dot{\bpsi_2}\trans S^{-1}\dot{\bpsi_2}-\dot{\bpsi_1}\trans A^{-1}\dot{\bpsi_1}\\
    &=\dot{\bpsi_1}\trans A^{-1}BS^{-1}CA^{-1}\dot{\bpsi_1}-\dot{\bpsi_1}\trans A^{-1}BS^{-1}\dot{\bpsi_2}-\dot{\bpsi_2}\trans S^{-1}CA^{-1}\dot{\bpsi_1}+\dot{\bpsi_2}\trans S^{-1}\dot{\bpsi_2}\\
    &=(S^{-1/2}CA^{-1}\dot{\bpsi_1}- S^{-1/2}\dot{\bpsi_2})\trans (S^{-1/2}CA^{-1}\dot{\bpsi_1}- S^{-1/2}\dot{\bpsi_2})\succeq 0.
\end{align*}
Therefore, we prove that QIF methods with more specified matrices can improve the efficiency. This proof can be easily generalized to more specified matrices if you treat $\bpsi_2$ as a larger estimating equation that consists of equations with different pre-specified matrices. 
\section{Additional results for simulation}
\subsection{Complete simulation results under various machine learning methods}
Tables \ref{con:20}-\ref{dur:1000} exhibit the simulation results for constant and duration-specific treatment effects under 
different machine learning methods, including
unadjusted, linear regression, regression tree, random forest (RF), XGBoost, SVM, neural network (NNet), and superlearner (SL) that uses an ensemble of regression trees, random forests, generalized linear models, and support vector machines. The summary results indicate that Superlearner, Regression Tree, and Random Forest consistently achieve the best overall performance. Furthermore, QIF can obtain additional efficiency gains. We also compare the performance of different machine learning methods with mixed-effect model. The linear mixed-effects model exhibits noticeable undercoverage in many settings. These conclusions further validate the properties of our estimator mentioned in the main paper.

\begin{table}[H]
		\centering
         \caption{Summary results for estimating the constant treatment effect with different machine learning methods on cluster-randomized trials with $I=20$ clusters.  We report the bias, empirical standard error (ESE), average of estimated standard errors (ASE), and coverage probability of the 95\% confidence interval (CP).}
        \label{con:20}
		\setlength{\tabcolsep}{5mm}{
        
			\begin{tabular}{cccccc}
				\toprule
				 Correlation  & Method  &  Bias       &  ESE  &  ASE   &  CP    
				\\ \midrule
				Independence&Unadjusted&-0.051&0.872&0.859&0.94\\
                & Linear & -0.011 & 0.614     & 0.547      & 0.942 \\
                    &Tree&-0.031 &0.537&0.463 &0.958 \\
                    &RF&-0.028 &0.539 &0.497 &0.954 \\
                    &XGBoost&-0.017 &0.486 &0.423 &0.951 \\
                    &SVM&-0.014 &0.552 &0.560 &0.968 \\
                    &NNet&-0.001 &0.714 &0.686 &0.964 \\
                    &SL& -0.024 & 0.474 &0.415  &0.952 
				\\ \midrule
				QIF &Unadjusted&0.021&0.313&0.327&0.954\\
                & Linear &0.017 & 0.271     & 0.263
                &  0.952 \\
                    &Tree&-0.018 &0.322 &0.294 &0.934 \\
                    &RF&0.010&0.215 &0.211 &0.95 \\
                    &XGBoost&-0.008 &0.241 &0.226 &0.945 \\
                    &SVM&-0.007 &0.275&0.283 &0.96 \\
                    &NNet&0.003 &0.494 &0.483 &0.939 \\
                    &SL& 0.002 &0.248&0.225  &0.936  \\ \midrule
                    Random effects&Linear&0.018  &0.252  &0.263  & 0.97
				\\ \bottomrule
		\end{tabular}}
	\end{table} 

\begin{table}[H]
		\centering
\caption{Summary results for estimating the constant treatment effect with different machine learning methods on cluster-randomized trials with $I=100$ clusters.  We report the bias, empirical standard error (ESE), average of estimated standard errors (ASE), and coverage probability of the 95\% confidence interval (CP).}
        \label{con:100}
		\setlength{\tabcolsep}{5mm}{
			\begin{tabular}{cccccc}
				\toprule
				 Correlation  & Method  &  Bias       &  ESE  &  ASE   &  CP    
				\\ \midrule
				Independence &Unadjusted&-0.016&0.436&0.431&0.956\\
                & Linear & 0.002 & 0.242      &  0.241      & 0.952  \\
                    &Tree&-0.001 &0.120&0.110 &0.96 \\
                    &RF&-0.005 &0.168 &0.157 &0.945 \\
                    &XGBoost&-0.003 &0.127 &0.125 &0.953 \\
                    &SVM&-0.002 &0.183 &0.199 &0.968\\
                    &NNet&-0.007 &0.203 &0.243 &0.976 \\
                    &SL& -0.003 & 0.123 & 0.112 & 0.959
				\\ \midrule
				QIF &Unadjusted&0.014&0.136&0.140&0.958\\
                & Linear &0.007 & 0.094     & 0.095
                &  0.95 \\
                    &Tree&0.004 &0.069 &0.068 &0.944 \\
                    &RF&0.006&0.062 &0.063 &0.945 \\
                    &XGBoost&0.002 &0.064 &0.063 &0.942 \\
                    &SVM&0.003 &0.085 &0.939 &0.971  \\
                    &NNet&-0.005 &0.154 &0.184 &0.978 \\
                    &SL& 0.006 &0.057& 0.057 & 0.944 \\ \midrule
                    Random effects&Linear  &0.007&0.124&0.129&0.963      
				\\ \bottomrule
		\end{tabular}}
	\end{table} 
 \begin{table}[H]
		\centering
\caption{Summary results for estimating the constant treatment effect with different machine learning methods on individual-randomized trials with $I=1000$ individuals.  We report the bias, empirical standard error (ESE), average of estimated standard errors (ASE), and coverage probability of the 95\% confidence interval (CP).}
\label{con:1000}
		\setlength{\tabcolsep}{5mm}{
			\begin{tabular}{cccccc}
				\toprule
				 Correlation  & Method  &  Bias       &  ESE  &  ASE   &  CP    
				\\ \midrule
				Independence &Unadjusted&0.006&0.347&0.353&0.952\\
                & Linear & 0.005 & 0.209     & 0.219      & 0.965 \\
                    &Tree&-0.001 &0.077& 0.075&0.962 \\
                    &RF&0.001 &0.118 &0.118 &0.953 \\
                    &XGBoost&0.001 &0.102 &0.096 &0.96 \\
                    &SVM&0.000 &0.142 &0.153 &0.973 \\
                    &NNet&-0.002 &0.110 & 0.114&0.96 \\
                    &SL& 0.000 &0.063  & 0.058 &0.961 
				\\ \midrule
				QIF &Unadjusted&0.014&0.181&0.177&0.946\\
                & Linear &0.003 & 0.109     & 0.112                & 0.952  \\
                    &Tree&0.017 &0.040 &0.040 &0.958 \\
                    &RF&0.004&0.058 &0.056 &0.948 \\
                    &XGBoost&0.001 &0.048 &0.044 &0.94 \\
                    &SVM&0.004 &0.081& 0.083&0.967 \\
                    &NNet&-0.002 &0.089 &0.085 &0.957 \\
                    &SL& 0.000 &0.037& 0.034 &0.934  \\
                    \midrule
                    Random effects&Linear& 0.010 &0.138  & 0.095 & 0.812

				\\ \bottomrule
		\end{tabular}}
	\end{table} 

 \begin{table}[H]
		\centering
\caption{Summary results for estimating the averaged duration-specific treatment effect with different machine learning methods on cluster-randomized trials with $I=20$ clusters.  We report the bias, empirical standard error (ESE), average of estimated standard errors (ASE), and coverage probability of the 95\% confidence interval (CP).}
\label{dur:20}
		\setlength{\tabcolsep}{5mm}{
			\begin{tabular}{cccccc}
				\toprule
				 Correlation  & Method  &  Bias       &  ESE  &  ASE   &  CP    
				\\ \midrule
				Independence 
                & Unadjusted & -0.037 & 1.081 & 1.085 & 0.934 \\
                & Linear &0.005  &0.712      &0.700       &0.947  \\
                    &Tree&-0.016 &0.572&0.483 &0.954 \\
                    &RF& -0.020&0.617 &0.581 &0.949 \\
                    &XGBoost&-0.015 &0.547 &0.509 &0.971 \\
                    &SVM&0.001 &0.646 &0.690 &0.963 \\
                    &NNet& -0.007&0.737 &0.732 &0.97 \\
                    &SL& -0.006 &0.518  &0.462  &0.957 
				\\ \midrule
				QIF & Unadjusted & -0.002 & 0.767 & 0.803 & 0.943 \\
                & Linear &-0.003 &0.581      &0.572 
                & 0.949  \\
                    &Tree&-0.015 &0.354 &0.347 &0.958 \\
                    &RF&0.016&0.420 &0.408 &0.942 \\
                    &XGBoost&-0.001 &0.357 &0.338 &0.96 \\
                    &SVM&0.016 &0.540&0.571 &0.965 \\
                    &NNet&0.026 &0.664&0.651 &0.96 \\
                    &SL& 0.010 &0.377& 0.328 &0.943  \\
                    \midrule
                    Random effects& Linear& -0.007 &0.552  &0.459  &0.911

				\\ \bottomrule
		\end{tabular}}
	\end{table} 

 \begin{table}[H]
		\centering
\caption{Summary results for estimating the averaged duration-specific treatment effect with different machine learning methods on cluster-randomized trials with $I=100$ clusters.  We report the bias, empirical standard error (ESE), average of estimated standard errors (ASE), and coverage probability of the 95\% confidence interval (CP).}
\label{dur:100}
		\setlength{\tabcolsep}{5mm}{
			\begin{tabular}{cccccc}
				\toprule
				 Correlation  & Method  &  Bias       &  ESE  &  ASE   &  CP    
				\\ \midrule
				Independence 
                & Unadjusted &  0.032 & 0.595 & 0.611 & 0.953 \\
                & Linear      &  0.021 & 0.361 & 0.360 & 0.942 \\
                    &Tree&0.000 &0.160&0.133 &0.973 \\
                    &RF& 0.009&0.198 &0.195 &0.946 \\
                    &XGBoost&-0.002 &0.156 &0.0.155 &0.98 \\
                    &SVM&0.019 &0.276 &0.298 &0.966 \\
                    &NNet& 0.017&0.295 &0.309 &0.976 \\
                    &SL&   0.004 & 0.143 & 0.124 & 0.96 
				\\ \midrule
				QIF & Unadjusted & -0.034 & 0.381 & 0.351 & 0.928 \\
                & Linear &  0.003 & 0.253 & 0.241 & 0.944 \\
                    &Tree&0.003 &0.071 &0.070 &0.95 \\
                    &RF&0.008&0.118 &0.110 &0.935 \\
                    &XGBoost&-0.019 &0.081 &0.085 &0.952 \\
                    &SVM&-0.002 &0.196&0.187 &0.945 \\
                    &NNet&0.013 &0.183&0.174 &0.961 \\
                    &SL&   0.006 & 0.071 & 0.069 & 0.939 \\
                    \midrule
                    Random effects& Linear& -0.015 & 0.309 & 0.213 & 0.824

				\\ \bottomrule
		\end{tabular}}
	\end{table}

 \begin{table}[H]
		\centering
\caption{Summary results for estimating the averaged duration-specific treatment effect with different machine learning methods on individual-randomized trials with $I=1000$ individuals.  We report the bias, empirical standard error (ESE), average of estimated standard errors (ASE), and coverage probability of the 95\% confidence interval (CP).}
\label{dur:1000}
		\setlength{\tabcolsep}{5mm}{
			\begin{tabular}{cccccc}
				\toprule
				 Correlation  & Method  &  Bias       &  ESE  &  ASE   &  CP  
				\\ \midrule
				Independence &Unadjusted&-0.014&0.479&0.484&0.945\\
                & Linear & -0.008 & 0.327     &0.327       & 0.949 \\
                    &Tree& -0.006&0.208&0.208 &0.952 \\
                    &RF&-0.009 &0.259 &0.263 &0.953 \\
                    &XGBoost&-0.006 &0.174 &0.176 &0.955 \\
                    &SVM&-0.004 &0.143 &0.149 &0.964 \\
                    &NNet&-0.009 &0.445 &0.454 &0.961 \\
                    &SL& -0.003 & 0.138 & 0.140 & 0.953
				\\ \midrule
				QIF &Unadjusted&-0.001&0.399&0.394&0.942\\
                & Linear     & -0.007 & 0.276 & 0.279 & 0.946 \\
                    &Tree&-0.007 &0.162 &0.163 &0.952 \\
                    &RF&-0.003&0.204 &0.204 &0.945 \\
                    &XGBoost&-0.004 &0.121 &0.118 &0.942 \\
                    &SVM& -0.003&0.090&0.093 &0.96 \\
                    &NNet&-0.016 &0.382 &0.328 &0.943 \\
                    &SL& -0.004 &0.077& 0.075 &0.954  \\
                    \midrule
                    Random effects        & Linear     &  0.044 & 0.295 & 0.116 & 0.558

				\\ \bottomrule
		\end{tabular}}
	\end{table}

\subsection{Simulation design for period-specific and saturated treatment effect structures.}
In this section, we present the simulation design for period-specific and saturated treatment effect structures. This simulation includes both cluster-randomized and individual-randomized trials.

For cluster-randomized trials, we generate 1{,}000 realizations under two trial sizes, with $I=20$ and $I=100$ clusters. When $I=20$, we set $J=3$, $N_i=20$, and $N_{ij}$ uniformly from $[5,15]$. When $I=100$, we set $J=5$, $N_i=500$, and sample $N_{ij}$ uniformly from $[5,35]$. We independently generate four covariates: one cluster-level covariate $X_{i1}\sim N(0,1)$ and three individual-level covariates, $X_{ik2}\sim \mathrm{Bernoulli}(0.5)$ and $X_{ik3}, X_{ik4}\sim N(0,1)$. Sampling indicators $S_{ijk}$ follows uniform distribution given $N_{ij}$ and $N_i$. Treatment assignments $Z_i$ are generated independently across clusters, with equal probability to start treatment in periods $1,\dots, J$. 

Under the period-specific setting, outcomes are generated as
\begin{align*}
    Y_{ijk} &= I\{Z_i \le j\}\,\beta_j(X) + \exp(X_{ik1}X_{ik2})
+ \tfrac{1}{2}X_{ik4}^2 + I\{X_{ik4}>-1\}+ 2I\{X_{ik3}>1\} \notag \\
&\quad + I\{X_{ik1}>0.5\}(j+1) + \sigma_i + \alpha_{ij} + \tau_{ik} + \epsilon_{ijk},
\end{align*}
where $\sigma_i,\alpha_{ij}, \tau_{ik} \sim N(0,0.1)$, $\epsilon_{ijk} \sim N(0,0.7)$ and $\beta_j(X) = 1 + \bigl(X_{ik3}-\overline{X}_{i\cdot 3}\bigr)*j/2 + \bigl(X_{ik4}^3-\overline{X_{i\cdot 4}^3}\bigr)*j/J$. Under the saturated structure, outcomes are generated as
\begin{align*}
    Y_{ijk} &= \beta_{jd}(X) + \exp(X_{ik1}X_{ik2})
+ X_{ik4} + I\{X_{ik4}>-1\}+ I\{X_{ik3}>1\} \notag \\
&\quad + I\{X_{ik1}>0.5\}(j+1)/2 + \sigma_i + \alpha_{ij} + \tau_{ik} + \epsilon_{ijk},
\end{align*}
with $\beta_{jd}(X) = I\{d>0\}\{1+(X_{ik3}-\overline{X_{i.3}})(J-j)/4+(X_{ik4}^3-\overline{X_{i.4}^3})*d/8\}$ for $d=j-Z_i+1$. 

For individual randomized trials, we generate 1{,}000 datasets setting $I=1000$ with $J=20$. By design, $N_i=1$, and we set $N_{ij} \sim \textup{Bernoulli}(0.5)$. Covariates and treatments are generated as in cluster-randomized trials. 

Under the period-specific setting, outcomes are generated as
\begin{align*}
    Y_{ijk} &= I\{Z_i \le j\}\,\beta_j'(X) + \exp(X_{ik1}X_{ik2})/2
+ \tfrac{1}{2}X_{ik4}^2 + I\{X_{ik4}>-1\}+ 2I\{X_{ik3}>1\} \notag \\
&\quad + I\{X_{ik1}>0.5\}(j+1)  +\tau_{ik} + \epsilon_{ijk},
\end{align*}
where $\tau_{ik} \sim N(0,0.1)$, $\epsilon_{ijk} \sim N(0,0.9)$ and $\beta_j'(X) = 1 + \bigl(X_{ik3}-\overline{X}_{i\cdot 3}\bigr)*j/J + \bigl(X_{ik4}^3-\overline{X_{i\cdot 4}^3}\bigr)/J$. Under the saturated structure, outcomes are generated as
\[
Y_{ijk}=\beta_{jd}'(X)+\frac{1}{2}X_{ik3}^3+\frac{1}{2}X_{ik4}^3+\frac{J+1}{2}X_{ik1} X_{ik2}+I\{X_{ik4}>0.5\}+\tau_{ik}+\epsilon_{ijk}
\]
with $\beta_{jd}'(X) = I\{d>0\}\{1+(X_{ik3}-\overline{X_{i.3}})*j+(X_{ik4}^3-\overline{X_{i.4}^3})*d\}$ for $d=j-Z_i+1$.

Tables \ref{per:20}-\ref{sat:1000} present the simulation results for estimating the average period-specific and saturated treatment effects using different machine learning methods and  mixed-effect models. Across all settings, our proposed estimators show negligible bias and achieve close-to-nominal 95\% coverage. Furthermore, covariate adjustment, machine learning–based modeling, and QIF all lead to efficiency gains. These results further support our asymptotic theory. Linear mixed models exhibit noticeable undercoverage in the large-sample setting, which is consistent with what we report in the main paper.

 \begin{table}[H]
		\centering
\caption{Summary results for estimating the averaged period-specific treatment effect with different machine learning methods on cluster-randomized trials with $I=20$ clusters.  We report the bias, empirical standard error (ESE), average of estimated standard errors (ASE), and coverage probability of the 95\% confidence interval (CP).}
\label{per:20}
		\setlength{\tabcolsep}{5mm}{
			\begin{tabular}{cccccc}
				\toprule
				 Correlation  & Method  &  Bias       &  ESE  &  ASE   &  CP    
				\\ \midrule
			Independence &Unadjusted&-0.022 &0.663 & 0.656&0.932\\
            & Linear & -0.003 & 0.445     & 0.440      & 0.965 \\
                    &Tree&-0.021 &0.380&0.349 &0.95 \\
                    &RF&-0.015 &0.388 &0.367 & 0.961\\
                    &XGBoost&-0.012 &0.362 &0.366 &0.969 \\
                    &SVM&-0.006 &0.408 &0.436 &0.966 \\
                    &NNet&-0.011 &0.515 &0.534 &0.964 \\
                    &SL& -0.012 & 0.341 & 0.323 & 0.953\\
                    &LM& -0.005 &0.430  & 0.193 &0.644
				\\ \midrule
				QIF& Unadjusted&-0.002 &0.379 &0.364 &0.924\\
                & Linear &-0.009 & 0.321     & 0.293
                &  0.935 \\
                    &Tree&-0.018 &0.294 &0.253 &0.93 \\
                    &RF&0.015&0.258 &0.246 &0.947 \\
                    &XGBoost&-0.017 &0.254 &0.249 &0.941 \\
                    &SVM&-0.011 &0.305& 0.298&0.946 \\
                    &NNet&-0.002 &0.466 &0.458 &0.945 \\
                    &SL& 0.001 &0.255& 0.245 &0.938  \\
                    \midrule
                    Random effects&Linear& 0.006 & 0.307 &0.260  & 0.919

				\\ \bottomrule
		\end{tabular}}
	\end{table} 
   \begin{table}[H]
		\centering
\caption{Summary results for estimating the averaged period-specific treatment effect with different machine learning methods on cluster-randomized trials with $I=100$ clusters.  We report the bias, empirical standard error (ESE), average of estimated standard errors (ASE), and coverage probability of the 95\% confidence interval (CP).}
\label{per:100}
		\setlength{\tabcolsep}{5mm}{
			\begin{tabular}{cccccc}
				\toprule
				 Correlation  & Method  &  Bias       &  ESE  &  ASE   &  CP    
				\\ \midrule
				Independence &Unadjusted&0.022&0.531&0.532&0.951\\
                & Linear & 0.011 & 0.235     & 0.235      & 0.958 \\
                    &Tree&0.004 &0.248&0.216 &0.954 \\
                    &RF&0.006 &0.279 &0.257 &0.967 \\
                    &XGBoost&0.006 &0.276 &0.262 &0.968 \\
                    &SVM& 0.004&0.243 &0.262 &0.975 \\
                    &NNet&0.007 &0.251 &0.273 &0.977 \\
                    &SL& 0.006 &0.205  &0.177  & 0.95
				\\ \midrule
				QIF &Unadjusted&0.017&0.175&0.170&0.952\\
                & Linear &0.002 & 0.108     & 0.108
                &  0.95 \\
                    &Tree&-0.000 &0.111 &0.108 &0.942 \\
                    &RF&0.009&0.085 &0.084 &0.962 \\
                    &XGBoost&0.008 &0.090 &0.089 &0.959 \\
                    &SVM&0.002 &0.106&0.115 &0.973 \\
                    &NNet&0.003 &0.144 &0.153 &0.977 \\
                    &SL& 0.003 &0.062& 0.062 &0.95  \\ \midrule
                    Random effects&Linear&0.001&0.139&0.117&0.899

				\\ \bottomrule
		\end{tabular}}
	\end{table} 
  \begin{table}[H]
		\centering
\caption{Summary results for estimating the averaged period-specific treatment effect with different machine learning methods on individual-randomized trials with $I=20$ individuals.  We report the bias, empirical standard error (ESE), average of estimated standard errors (ASE), and coverage probability of the 95\% confidence interval (CP).}
\label{per:1000}
		\setlength{\tabcolsep}{5mm}{
			\begin{tabular}{cccccc}
				\toprule
				 Correlation  & Method  &  Bias       &  ESE  &  ASE   &  CP    
				\\ \midrule
				Independence &Unadjusted&0.004 &0.348 & 0.349&0.953\\
                & Linear &0.002  & 0.212     & 0.221      & 0.965 \\
                    &Tree&-0.001 &0.080& 0.080& 0.954\\
                    &RF&0.001 &0.109 &0.109 &0.951 \\
                    &XGBoost&0.001 &0.084 &0.085 &0.959 \\
                    &SVM& 0.001& 0.149&0.158 &0.966 \\
                    &NNet&0.002 &0.111 &0.110 &0.956 \\
                    &SL& -0.001 & 0.069 & 0.068 &0.948 
				\\ \midrule
				QIF &Unadjusted&0.005 &0.156 &0.153 &0.946\\
                & Linear &-0.002 &0.101      &0.101 
                &  0.955 \\
                    &Tree& 0.001&0.042 &0.043 &0.952 \\
                    &RF&0.001&0.050 &0.050 &0.944 \\
                    &XGBoost&0.003 &0.036 &0.034 &0.937 \\
                    &SVM&0.001 &0.077&0.076 &0.938 \\
                    &NNet&-0.004 &0.073 &0.072 &0.949 \\
                    &SL& 0.001 &0.028& 0.027 & 0.957 \\
                    \midrule
                Random effects&Linear& 0.012 & 0.186 & 0.100 & 0.713

				\\ \bottomrule
		\end{tabular}}
	\end{table} 

\begin{table}[H]
		\centering
\caption{Summary results for estimating the averaged saturated treatment effect with different machine learning methods on cluster-randomized trials with $I=20$ clusters.  We report the bias, empirical standard error (ESE), average of estimated standard errors (ASE), and coverage probability of the 95\% confidence interval (CP).}
\label{sat:20}
		\setlength{\tabcolsep}{5mm}{
			\begin{tabular}{cccccc}
				\toprule
				 Correlation  & Method  &  Bias       &  ESE  &  ASE   &  CP    
				\\ \midrule
				Independence &Unadjusted&-0.024 &0.696 &0.637 &0.951\\
                & Linear &-0.006  &0.442      &0.384       &0.941  \\
                    &Tree&-0.027 &0.473&0.391 &0.953 \\
                    &RF& -0.018& 0.471&0.387 &0.958 \\
                    &XGBoost&-0.019 &0.432 &0.369 &0.978 \\
                    &SVM&-0.014 &0.428 &0.409 &0.967 \\
                    &NNet&-0.015 &0.507 &0.466 &0.96 \\
                    &SL&-0.013  &0.395  &0.305  &0.95 
				\\ \midrule
				QIF &Unadjusted&0.005 &0.553 &0.592 &0.943\\
                & Linear &0.011 &0.346      & 0.310
                &  0.949 \\
                    &Tree&-0.006 &0.336 &0.312 &0.949 \\
                    &RF&0.013&0.305 &0.287 &0.951 \\
                    &XGBoost&-0.014 &0.271 &0.269 &0.967 \\
                    &SVM&0.009 &0.312&0.304 &0.963 \\
                    &NNet&0.012 &0.398 &0.386 &0.97 \\
                    &SL& 0.008 &0.244& 0.218 &0.954  \\
                    \midrule
                    Random effects&Linear& 0.009 &0.308  &0.288  & 0.944

				\\ \bottomrule
		\end{tabular}}
	\end{table} 
      \begin{table}[H]
		\centering
\caption{Summary results for estimating the averaged saturated treatment effect with different machine learning methods on cluster-randomized trials with $I=100$ clusters.  We report the bias, empirical standard error (ESE), average of estimated standard errors (ASE), and coverage probability of the 95\% confidence interval (CP).}
\label{sat:100}
		\setlength{\tabcolsep}{5mm}{
			\begin{tabular}{cccccc}
				\toprule
				 Correlation  & Method  &  Bias       &  ESE  &  ASE   &  CP    
				\\ \midrule
				Independence &Unadjusted&0.009 &0.267 &0.257 &0.949\\
                & Linear &0.007  &   0.197   & 0.190      & 0.942 \\
                    &Tree&0.002 &0.135&0.121 &0.959 \\
                    &RF&0.004 &0.157 &0.145 &0.953 \\
                    &XGBoost&0.003&0.165 &0.156 & 0.953 \\
                    &SVM&0.001 &0.149 &0.156 &0.978 \\
                    &NNet&0.005 &0.186 &0.223 &0.981 \\
                    &SL& 0.000 & 0.124 & 0.105 & 0.951
				\\ \midrule
				QIF &Unadjusted&-0.006 &0.158 &0.150 &0.934\\
                & Linear &0.001 & 0.145     & 0.136
                &  0.939 \\
                    &Tree&0.001 &0.087 &0.088 &0.961 \\
                    &RF&0.012&0.094 &0.095 &0.948 \\
                    &XGBoost&-0.026 &0.095 &0.098 &0.953 \\
                    &SVM&0.005 &0.091&0.098 &0.968 \\
                    &NNet&0.007 &0.197 &0.145 &0.972 \\
                    &SL& 0.002 &0.063& 0.064 &0.954  \\ \midrule
                   
                    Random effects&Linear&-0.003&0.132&0.105&0.89

				\\ \bottomrule
		\end{tabular}}
	\end{table} 

      \begin{table}[H]
		\centering
\caption{Summary results for estimating the averaged saturated treatment effect with different machine learning methods on individual-randomized trials with $I=1000$ individuals.  We report the bias, empirical standard error (ESE), average of estimated standard errors (ASE), and coverage probability of the 95\% confidence interval (CP).}
\label{sat:1000}
		\setlength{\tabcolsep}{5mm}{
			\begin{tabular}{cccccc}
				\toprule
				 Correlation  & Method  &  Bias       &  ESE  &  MSE   &  CP    
				\\ \midrule
				Independence &Unadjusted&-0.005&0.390&0.381&0.948\\
                & Linear &0.001&0.258&0.260&0.947  \\
                    &Tree&-0.009&0.170&0.164&0.946 \\
                    &RF&-0.006&0.201&0.197&0.952 \\
                    &XGBoost&-0.003 &0.129 &0.122 &0.943 \\
                    &SVM&-0.002 &0.099 &0.099 &0.957 \\
                    &NNet& 0.021& 0.462&0.484 &0.977 \\
                    &SL& -0.002 &0.091  &0.089  &0.96
				\\ \midrule
				QIF &Unadjusted&0.007&0.142&0.145&0.957\\
                & Linear &-0.001&0.123&0.121&0.943 \\
                    &Tree&-0.012&0.093&0.096&0.946 \\
                    &RF&0.002&0.067&0.067&0.953 \\
                    &XGBoost& -0.009& 0.062&0.061 &0.938 \\
                    &SVM&-0.007 &0.064&0.066 &0.947 \\
                    &NNet&-0.015 &0.282 &0.268 &0.932 \\
                    &SL& 0.001 &0.044& 0.044 & 0.947 \\
                    \midrule
                    Random effects&Linear& 0.008 &0.208  &0.088  & 0.607

				\\ \bottomrule
		\end{tabular}}
	\end{table} 

\section{Imbalance of baseline covariates of the second real data.}
\begin{figure}[H]  
    \centering
    \includegraphics[width=0.8\textwidth]{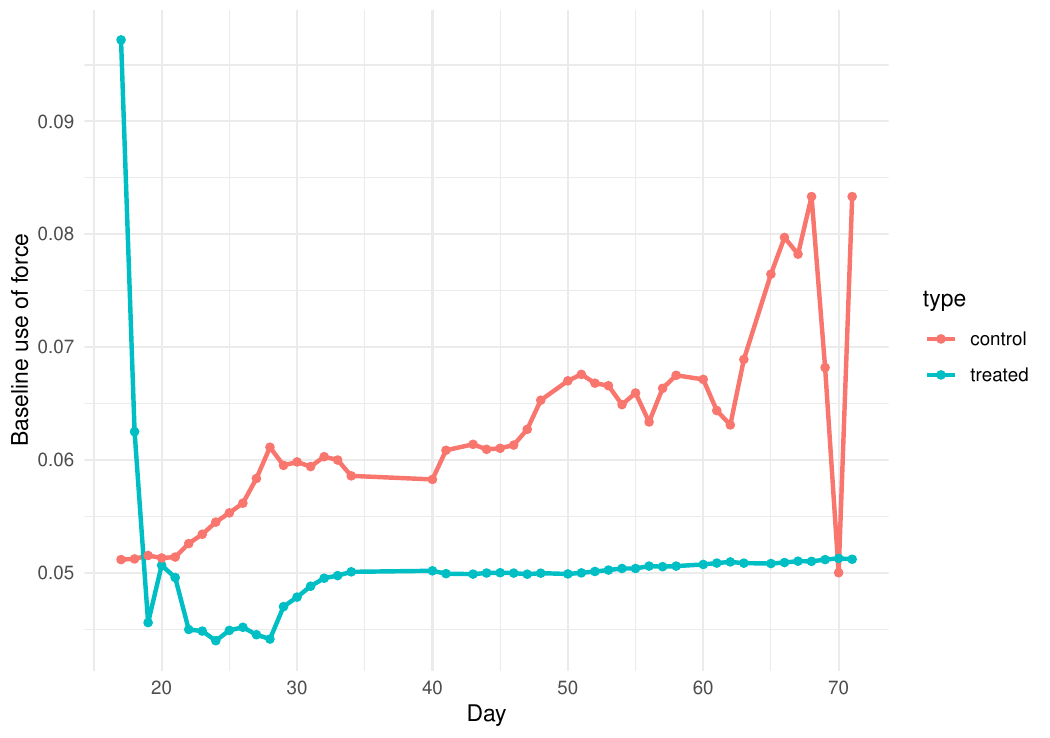} 
    \caption{Imbalance of baseline use of force in the Chicago training program.}
    \label{fig:1}
\end{figure}

Figure \ref{fig:1} exhibits the imbalance of baseline covariates, use of force among the treatment and control groups. This figure indicates that officers in training have a lower baseline use of force compared to those in the control group.

\bibliographystyle{apalike}
\bibliography{library}